# Gene regulatory interactions limit the gene expression diversity

Orr Levy[1,2], Shubham Tripathi[3], Scott D. Pope[1,2], Yang-Yu Liu[4,5,*] and Ruslan Medzhitov[1,2,6,7,*]

[1]Department of Immunobiology, Yale University School of Medicine, New Haven, CT, USA

[2]Howard Hughes Medical Institute, Chevy Chase, MD, USA

[3]Yale Center for Systems and Engineering Immunology and Department of Immunobiology, Yale School of Medicine, New Haven, CT 06520, USA

[4]Channing Division of Network Medicine, Brigham and Women's Hospital and Harvard Medical School, Boston, Massachusetts 02115, USA.

[5]Center for Artificial Intelligence and Modeling, The Carl R. Woese Institute for Genomic Biology, University of Illinois at Urbana-Champaign, Urbana, IL 61801, USA.

[6]Tananbaum Center for Theoretical and Analytical Human Biology, Yale University School of Medicine, New Haven, CT, USA

[7]Lead Contact

*Correspondence: ruslan.medzhitov@yale.edu ; yyl@channing.harvard.edu

## Abstract

**The diversity of gene expression plays a critical role in cellular specialization, adaptation to environmental changes, and overall cell functionality. This diversity varies dramatically across different cell types and is orchestrated by intricate, dynamic, and cell type-specific gene regulatory networks (GRNs). Despite extensive research on GRNs, the governing principles that have shaped them remain largely unknown. Here we investigated whether there is a tradeoff between the diversity of gene expression and the intensity of gene regulations. Leveraging a computational method that evaluates GRN interaction intensity from scRNA-seq data, we analyzed both simulated data and real scRNA-seq data collected from different tissues in humans, mice, fruit flies, and *C. elegans*. We found a significant tradeoff between diversity and interaction intensity, driven by stability constraints, where the GRN could be stable up to a critical level of complexity - a product of gene expression diversity and interaction intensity. Furthermore, we analyzed hematopoietic stem cell differentiation data and find that the overall complexity of unstable transition states cells is higher than that of stem cells and fully differentiated cells. Our results suggest that GRNs are shaped by stability constraints which limit the diversity of gene expression.**

## Introduction

The diversity of expressed genes in different cell types is critical for cellular specialization, adaptation to environmental changes, and overall cell function. This diversity is inherently intertwined with the distinctive functional properties of different cell types. While some cell types exhibit a wide array of functions, others have a more limited functional repertoire. For example, hepatocytes, the primary cells in the liver, have diverse metabolic functions, including detoxification, synthesis of plasma proteins, lipid metabolism, and regulation of glucose homeostasis [1]. In contrast, muscle cells are an example of a highly specialized cell type [1,2], having a narrow range of functions and a relatively low number of expressed genes. The functional diversity inherent to each cell type intricately shapes its unique gene expression profile. Each cellular function is orchestrated by a dedicated gene program comprising a specific ensemble of genes. Consequently, distinct tissues and cell types exhibit substantial disparities in the diversity of genes they express [3,4].

These gene expression programs are governed by complex, dynamic, and cell-type-specific Gene Regulatory Networks (GRNs) [5], which optimize the cell’s functionality [6] and maintain its stability [7]. GRNs consist of "nodes," which include genes and their regulators, connected by "edges" that depict regulatory interactions. Extensive studies of GRNs found structural features of complex GRNs, having a variety of network motifs [8], power-law outdegree distribution [9], and an exponential indegree distribution [10]. However, the global characteristics of GRNs remain largely unknown [11]. Their governing principles that ensure appropriate gene expression programs and the underlying driving forces that shaped them remain elusive [12].

Here, we suggest a hypothesis that there is a tradeoff between the gene expression diversity and the GRN interaction intensity. Here we define interaction intensity as a function of network density and the strength of regulatory interactions. According to the hypothesis, it would be impossible for a cell, to have both a high level of GRN interaction intensity (Q) and a high gene expression diversity, suggesting that GRNs are bounded by an upper level of complexity, a product of the diversity of expressed genes and the interaction intensity (Fig. 1).

Our goal is to better understand the limitations of the gene expression diversity in GRNs. While the diversity of gene expression is, in principle, an observable measure [14,15], the GRNs are generally unknown. Here, we leverage a computational method that evaluates the level of interaction intensity of the GRN, from a cohort of data samples (scRNA-seq cells from a given cell type), without reconstructing the GRN. In this method, we assess the transcriptional multivariate interdependence among genes, denoted as the global coordination level (GCL). First, we demonstrated the GCL's ability to measure the underlying GRN's interaction intensity in simulated and real gene expression data. Using a classical GRN dynamical model [13], we simulated gene expression data and demonstrated the ability of the GCL to estimate the predetermined GRN's interaction intensity from the gene expression profile of the simulated cells. Then, we analyzed real scRNA-seq datasets and demonstrated the ability of the GCL to estimate increased levels of interaction intensity in known pathways. Second, using the GRN dynamical model, we simulated different cell types with different GRNs, each with a predetermined interaction intensity, finding a tradeoff between the diversity of expressed genes and the interaction intensity, a tradeoff that we also find when calculating the GCL on the simulated steady states. Third, we applied the GCL to analyze samples collected from different cell types and tissues, in human, mice, fruit fly and *C. elegans*, finding a significant tradeoff between the diversity of expressed genes and their interaction intensity. Finally, we analyzed scRNA-seq data from two distinct biological processes: differentiation, and stimulation. In marrow and peripheral blood differentiation scRNA-seq data, we found that during the transition from stem to differentiated states, cells exhibit significantly higher complexity than stem cells or fully differentiated cells. In macrophages, with distinct scRNA-seq gene expression profiles upon LPS and IL4 stimulation, we also found a significant tradeoff between the diversity of expressed genes and interaction intensity. Using numerical simulations and the real scRNA-seq data analysis, we showed that stability constraint is the main driver for the interaction intensity – diversity of gene expression tradeoff. These results suggest that cellular GRNs are shaped by stability constraints, which limit the interaction intensity and the diversity of expressed genes to an optimal level for their function.

## Results

### Evaluating gene-to-gene regulatory network interaction intensity with transcriptional coordination measure

Evaluating the intricacy of gene-to-gene regulatory interactions within a GRN poses a formidable challenge, even when considering its most elementary form, which involves interactions between Transcription Factors (TFs) and their target genes. In this simplified representation, the nodes within the GRN encompass various genes, including some functioning as TFs, while the network's edges symbolize the regulatory connections governing gene expression. Reconstructing Gene Regulatory Networks (GRNs) became a focus by utilizing diverse high-throughput experimental techniques and computational algorithms [16,17]. Without prior knowledge of the complex network of gene-to-gene regulatory interactions, various methods are frequently employed to elucidate the transcriptional relationships between genes to reconstruct the GRN. One common approach involves analyzing transcriptional co-expression patterns. In essence, the examination of alterations in co-expression networks calculated from single-cell data can be leveraged to reconstruct the GRN [18,19,20,21] and thereby estimate the intensity of their interactions.

Nevertheless, this "bottom-up" approach faces significant challenges. First, co-expression networks primarily capture the direct or indirect correlations between pairs of genes, even though a single gene may be controlled by multiple regulators [8,22,23]. Second, each co-expression measure is tailored to capture specific features, and they may not be optimal for depicting all types of gene-to-gene transcriptional relationships. For example, Pearson correlations primarily represent linear relationships. Third, extensively computed co-expression matrices often contain significant noise, introducing an additional challenge when attempting to differentiate them across various cohorts.

To address these challenges and to evaluate the global gene-to-gene regulatory network interaction intensity, we use a 'top-down' computational approach for assessing the system-wide transcriptional multivariate interdependence among genes, denoted as the global coordination level (GCL) [24,25]. This approach does not require inferring an entire network of pairwise correlations. The fundamental concept behind GCL analysis lies in quantifying the average multivariate interdependence among randomly selected subsets of genes. Specifically, we employ the "bias-corrected distance correlation" measure [26] (see Methods).

To test the relations between the GCL and the underlying GRN interaction intensity, we first applied the GCL analysis to simulated scRNA-seq data generated using the Co-regulation GRN dynamics model (See methods). We simulated cohorts of cells in steady state; each cohort simulates a different cell type, originate from a different GRN, with similar network topology (a random network is presented in Fig. 2), and a unique combination of network density $C$ and the interaction weights $W$. Subsequently, using the steady states and a linear regression process, we showed that the GCL positively correlates to the predetermined interaction intensity of the underlying GRN dynamics. We also tested the correlation between GCL and network interaction intensity for the simulated dynamics of three known real biological GRNs which have been shown to capture biological behavior across contexts: an epithelial-mesenchymal transition network [27], a network involved in human embryonic stem cell differentiation [27], and a GRN inferred previously from the gene expression profiles of breast tumor samples [28]. In all cases, we found that the GCL increases with an increase in the network interaction intensity (SI Fig.1 A).

To further demonstrate that increased GCL reflects elevated levels of underlying gene-to-gene interactions, we analyzed the GCL of functionally related gene sets (170 pathways according to the Kyoto Encyclopedia of Genes and Genomes (KEGG) database [29]) in different cell types from mice. We conducted a comparative analysis by assessing the GCL value of each pathway in relation to the GCL values of corresponding 'surrogate' gene sets. These surrogate gene sets were randomly assembled from a subset of genes that exhibit similar expression levels to the genes within the pathway under investigation. Consequently, each surrogate gene set preserved the original pathway's size and mirrored its expression profile while not representing any known KEGG pathway (refer to the Methods for a comprehensive explanation of the surrogate procedure). To evaluate whether the examined pathways tend to display enhanced coordination, we analyzed the distribution of GCL values for each pathway and compared it to the distribution observed in the surrogate gene sets. As the GCL values of the surrogate gene sets linked to the same pathway roughly conformed to a normal distribution, we computed a Z-score for each pathway concerning these surrogate gene sets. In the absence of any additional coordination among genes within the same pathway (the null expectation), only approximately 16% of the pathways in each cell type would be anticipated to have a Z-score exceeding 1, similar to what is observed in the surrogate gene sets. As demonstrated in SI Fig.2, a significantly larger proportion of pathways in each cell type indeed exhibit Z-scores exceeding 1. Previous research has already established cross-dataset consistency for elevated GCL levels in pathways, as evidenced in [24]. Finally, additional support demonstrating that significant GCL reflects elevated levels of underlying gene-to-gene interactions was found in a recent analysis [30] of multi-omics data, where both single-cell DNA methylation (scDNAm) and scRNA-seq data were analyzed. It was found that co-regulated gene set with coordinated methylation patterns, also have significant GCL.

## GRN intensity – diversity tradeoff is observed in simulated and real gene expression data.

We investigated the tradeoff between GRN interaction intensity and diversity of expressed genes (GRN intensity-diversity tradeoff) using a co-regulation GRN dynamics model, which represents GRN formed under stability constraints. That is, we only keep steady states of the simulated cells (Methods). Using the model, we generated a set of cohorts of cells, each with a unique set of $W$ and $C$, for each cohort of steady-state cells, we calculated the diversity of expressed genes, which we define as the effective number of expressed genes ($N_{\mathrm{eff}}$), using Shannon diversity [13] (Methods). Estimating the diversity of expressed genes through scRNA-seq data presents a formidable challenge, primarily because of the substantial presence of low-expressed genes that are challenging to identify. To address this challenge, we conducted our analysis using various measures (Methods) and later showed that a broader concept of interaction intensity-diversity holds true. In Fig. 2d, we calculated the gene expression diversity. It is shown that the cohorts with high interaction intensity (high C or high W) can be stable only when the diversity of expressed genes is relatively small, while a large diversity of expressed genes is found only in cohorts with low interaction intensity. In Fig. 2e, we can also see that, using the GCL measured over the steady states of the different cell types, we obtained a similar tradeoff, showing that the tradeoff could also be seen with the GCL calculated over the systems' steady states rather than directly calculating the interaction intensity of their underlying GRN. To further expand the validity of our findings to various network types, we applied a comprehensive analysis of GRNs from broader classes with different characteristics extracted from real data and found consistent results. The same tradeoff between GCL and $N_{\text{eff}}$ is seen for the simulated steady states of biological GRNs taken from the literature (SI Fig.1 b-c). A similar tradeoff is seen in the case of random GRNs that exhibit the minimal frustration property, a characteristic of biological GRNs across contexts [13] (SI Fig.3). Thus, the GCL-$N_{\text{eff}}$ tradeoff is not restricted to the case of random GRNs but extends to GRNs with biological features.

Next, to test our hypothesis on real scRNA-seq data, we systematically applied our method to different organisms and tissue datasets of scRNA seq. We analyzed hundreds of cohorts of different cell types from different organisms and a variety of tissues. Human Tabula Sapiens [31], mice Tabula Muris [32], Human Gut cell atlas [33], fruit fly cell atlas [34] and *C. elegans* atlas [35] (see SI Table 1 for detailed description). To minimize the effects of potential confounding factors related to different experimental setups, bioinformatics pipelines, or cell-type-specific regulatory dynamics, we compare GCL and $N_{\text{eff}}$ values calculated for cell types from each organism and from the same study separately, and calculate them per cohort of cells from a given sample (e.g., in each dataset we have several cohorts of each cell type, coming from different sample, tissue, sex, age, and etc). In SI. Fig. 4, we showed the assessment of the interaction intensity of the GRNs of different cell types, using the GCL. It is shown that GCL values vary significantly across different tissues, and cell types. To verify that the GCL of a given cell type is indeed a characterizing quantity, we validated that the GCL values of a cell type, are consistent across different samples of the same cell type collected from the same dataset, for the same age and from the same tissue (Methods and SI. Fig. 5). Our results show that GCL varies across different cell types, suggesting that they originate from GRNs with different interaction intensity.

We next compared the cohorts from each dataset separately. Fig. 3 shows the GCL-$N_{\text{eff}}$ tradeoff, that cohorts of cell types with a high diversity of expressed genes have a low GCL. To rigorously assess the statistical significance of the observed tradeoff patterns, as depicted in Fig.2e-f and Fig.3, we employed a curve fitting approach using a curve of the form $\frac{\alpha}{1+e^{-\beta(x+\gamma)}}+\varepsilon$ (where $\alpha$, $\varepsilon$, $\beta$, and $\gamma$ are fitting parameters: $\alpha=\max(GCL)$, $\varepsilon=\min(GCL)$) to characterize the empirical upper limit of the data points (Methods). This model effectively characterizes the empirical upper boundary of our data points in both simulated and real scRNA-sec data, ensuring the reliability of our findings. These insightful observations provide strong support for the hypothesis proposing a fundamental tradeoff between the diversity of gene expression and the intensity of interactions.

To assess whether our findings might be attributed solely to disparities in gene expression distributions among various cell types rather than the interactions between them, we conducted an additional analysis on shuffled gene expression tables. These shuffled tables were designed to maintain the gene expression distribution for each gene within each cell type while nullifying the impact of inter-gene interactions. Notably, our results from this analysis demonstrated a consistently flat GCL across all instances, and we were unable to replicate the $N_{\text{eff}}$-GCL pattern (see SI Fig. 6). This compelling observation underscores the significance of inter-gene interactions as a driving force behind the patterns we've identified, reinforcing the robustness of our primary findings. Finally, we tested two additional measures for the diversity of expressed genes and found the trade-off to be robust (SI Fig.7-8). Overall, this emerging principle serves as a governing factor of gene regulatory networks, advancing our comprehension of the intricate dynamics at play in cellular processes.

## GRN interaction intensity – Diversity of expressed genes tradeoff is observed in differentiation and upon stimulation.

To further explore the observed pattern in natural scRNA-seq data obtained from various tissues and cell types, we expanded our analysis to two different biological processes, where we expect that the tradeoff may play a significant role in cell fate. First, we analyzed Hematopoietic stem cells (HSCs) differentiation data, collected from two independent studies of mice [32] and human [36], from bone marrow and peripheral blood. We calculated the $N_{\mathrm{eff}}$ and GCL of each cohort and estimated its underlying GRN complexity by multiplying its GCL and $N_{\mathrm{eff}}$. For the purpose of estimating the complexity, in order to give a similar weight to the GCL and $N_{\mathrm{eff}}$, we used normalized versions of both GCL and $N_{\mathrm{eff}}$, by having the same mean value. We divided the cohorts of cells to three groups according to their differentiation stage (see SI Table 2), non-differentiated stem cells, partially differentiated cells (represents cell types in transition states) and fully differentiated cell types. We found a similar pattern across the two datasets, where stem cells and fully differentiated cells having much lower complexity (see Fig.4a) and GCL (see SI Fig.9), compared with transition states cells. In SI Fig.9 we further found that along differentiation, $N_{\mathrm{eff}}$ reduces significantly. To extend our analysis of changes in GCL during a cell-fate transition, we used a GRN involved in the epithelial-mesenchymal transition [37] to simulate cell-fate switching from epithelial to mesenchymal state (SI Fig.10). Our analysis showed that the GCL was higher for cells in the transition state, compared to both epithelial and mesenchymal cells (SI Fig.10F), consistent with the higher GCL seen in the case of partially differentiated human and mouse HSCs. These results, where unstable transition state cells having higher complexity, support our simulation findings, that the diversity-interaction intensity tradeoff is driven by stability constraints, where high complexity cells having lower stability.

In addition to cellular differentiation in which cells change identity, we analyzed the effects of cellular stimulation. We stimulated bone marrow-derived macrophages for 2 hours with the bacterial cell wall component lipopolysaccharide, the Th2 cytokine IL-4, or left the macrophages unstimulated and performed single cell sequencing. Both stimulations cause large changes in the GRN without affecting cellular identity. We found that the tradeoff between GCL and $N_{\text{eff}}$ exists upon stimulation (FIG. 4B). Interestingly, IL-4 and LPS both cause a decrease in $N_{\text{eff}}$ and an increase in the GCL but maintain the same complexity in contrast to the cellular differentiation (SI FIG. 11). We hypothesize that the increase in GCL is due to the reorganization of the coregulator p300 in the nucleus. Before stimulation, p300 is interacting with many different transcription factors which is driving many different steady-state gene expression programs (tissue surveillance, detection of bacteria and viruses, cellular migration, division, etc.) but upon stimulation and the activation of stimuli specific transcription factors (NF-kB, IRF and AP-1 for LPS; Stat6 for IL-4) the GRN is significantly altered and skews the p300 to be associated with the stimuli specific gene expression program at the expense of the steady-state gene expression programs. This may cause a smaller number of more highly connected gene expression programs to be expressed.

## Discussion

Multiple factors may dictate the substantial variations in gene expression diversity across different cell types, including functional requirements of cells in distinct tissues, metabolic and other resource constraints, and the intricate dynamical processes within the underlying Gene Regulatory Networks (GRN). The observed trade-off between the diversity of expressed genes and the intensity of GRN interactions underscores the significance of the latter. Supported by numerical simulations and real data analysis, we find an unfeasible region, where there are no cell types (in steady state) that have both high interaction intensity and high gene expression diversity. This region becomes feasible only transiently, when involving simulated cells that have not reached their steady state, furthermore, during differentiation process transition state cells which are biologically unstable, show an increased level of complexity. Thus, we suggest that to maintain stability, GRNs have to evolve to either small and highly interacting or to large but weakly interacting gene regulatory programs. Additionally, we explored the potential association of the observed tradeoff with energetic constraints by examining the correlation between cell size, a parameter linked to cellular energetic capacities [38], and complexity. However, no significant correlation was identified in our analysis (SI Fig. 12). An additional driving force to the observed tradeoff could be competitive exclusion [39], especially in cases of a significant fraction of repressive regulatory interactions. We noticed that the tradeoff also occurs in GRN models with strictly activating interactions supporting the stability constraint as a key driver.

The interaction intensity–diversity tradeoff, driven by stability constraints, was initially proposed by Robert May. His theoretical framework posits that the stability of a large-scale ecosystem, encompassing diverse species interacting randomly, is intricately tied to the complexity of interspecific interactions around an equilibrium point [40]. Subsequently, a more recent study, identified a comparable tradeoff within a specific class of complex dynamical networks [41]. Additionally, this tradeoff has been explored in a model of gene regulatory networks, suggesting an association between network topology and the stability of the GRN [7], yet this study is the first evidence of this tradeoff in real scRNA-seq data. It is important to state the assumptions we made to show the complexity–stability relationship using our methodology. Firstly, our model specifically concentrated on simulating cells within a designated class (e.g., network topology, degree distribution), whether random networks or actual GRNs with heightened interaction intensity; notably, these classes were not intermingled. Another assumption is that each set of individual cells within a particular cell type originates from the same GRN, and diverse cell types emerge from the same class of GRNs.

The governing principles that underlie the design and optimality of GRNs are not only crucial for advancing our fundamental understanding of cells but also hold significant implications for applications. For example, in tissue engineering [42], or therapeutic interventions [43], cells' environment changes dramatically and cells are targeted to reach a desired fate, which may lead to outcomes diverging from initial expectations due to the intricate interplay of complexity and stability considerations. Additionally, we notice that some cell types have higher complexity, suggesting they are at the boundary of stability.

Lastly, the approach for assessing the interaction intensity of GRNs can be valuable in examining the complexity–stability relationship and extending its utility to broader applications. In stable systems, the interaction intensity becomes pivotal in gauging the repercussions of gene perturbations across the entire GRN. Moreover, the established framework offers a means to evaluate the impact of perturbations on GRNs and their cascading effects on other genes. Additionally, it serves as a foundation for assessing individual genes and gene programs crucial to GRNs, eliminating the need to extract the entire network for analysis.

## Methods

**Bias corrected distance correlation (bcdCorr).** The level of dependence between two variables can be assessed using bcdCorr [26], an enhanced adaptation of the distance correlation (dCorr) [44], which is not affected by the vectors size. In essence, the distance correlation evaluates the extent of dependence between two variables by examining how the distances between pairs of samples, as measured by one variable, change in relation to the distances between the same pairs of samples, as measured by the other variable. In simpler terms, a strong dependency indicates that slight alterations in one variable correspond to minor adjustments in the other variable. This metric is adept at capturing non-linear relationships that might escape detection through conventional correlation techniques.

For instance, the correlation between two quadratically related variables, e.g., $y = x^2$ where x is symmetrically distributed around zero, is zero. However, when comparing two samples, i.e., to data points $(x_I, y_i)$ and $(x_j, y_j)$, it is obvious that the difference between the two samples in one dimension $(|x_I - x_j|)$ is correlated with the difference in the second dimension $(|y_I - y_j|)$. Similarly, the distance correlation can also be applied to measure the dependency between two high-dimensional variables. Furthermore, the utility of distance correlation extends to gauging the dependency between two variables in high-dimensional spaces. This analytical approach remains applicable and insightful in scenarios involving intricate relationships across multiple dimensions.

Consider $M$ observations of two high-dimensional variables $\boldsymbol{X_i} \in \mathbb{R}^p$ and $\boldsymbol{Y_i} \in \mathbb{R}^q, \quad i = 1, \dots, M$, where $\boldsymbol{X_i} = \left(X_{i,1}, \dots., X_{i,p}\right)$ and $\boldsymbol{Y_i} = \left(Y_{i,1}, \dots, Y_{i,q}\right)$. Note that $q$ does not necessarily have to be equal to $p$. The $M$ observations are represented by the $p \times M$ matrix $\boldsymbol{X}$ and $q \times M$ matrix $\boldsymbol{Y}$. The empirical $\mathrm{bcdCorr}(\boldsymbol{X}, \boldsymbol{Y})$ is defined as

$$\text{bcdCorr}\,(\boldsymbol{X},\boldsymbol{Y}) = \frac{\text{dCov}\,(\boldsymbol{X},\boldsymbol{Y})}{\sqrt{\text{dCov}\,(\boldsymbol{X},\boldsymbol{X})\cdot\text{dCov}\,(\boldsymbol{Y},\boldsymbol{Y})}}$$

where

$$\text{dCov}\,(\boldsymbol{X},\boldsymbol{Y}) = \frac{1}{M(M-3)}\left[\sum_{i,j=1}^{M} A^*_{i,j}B^*_{i,j} - \frac{M}{M-2}\sum_{i=1}^{M} A^*_{i,i}B^*_{i,i}\right]$$

and $A^*_{i,j}$ and $B^*_{i,j}$ are matrices defined as

$$A^*_{ij} = \begin{cases}\frac{M}{M-1}\left(A_{i,j} - \frac{a_{ij}}{M}\right), & i \neq j\\ \frac{M}{M-1}(\bar{a}_i - \bar{a}), & i = j\end{cases}, \qquad B^*_{ij} = \begin{cases}\frac{M}{M-1}\left(B_{i,j} - \frac{b_{ij}}{M}\right), & i \neq j\\ \frac{M}{M-1}(\bar{b}_i - \bar{b}), & i = j\end{cases}$$

and $A_{i,j}$ and $B_{i,j}$ are matrices defined as

$$A_{i,j} = a_{ij} - \bar{a}_i - \bar{a}_j + \bar{a},$$

$$B_{i,j} = b_{ij} - \bar{b}_i - \bar{b}_j + \bar{b},$$

where

$$a_{ij} = \left|\boldsymbol{X}_i - \boldsymbol{X}_j\right|, \quad i,j = 1,\dots,M,$$

$$a_{i\cdot} = \sum_{k=1}^{M} a_{ik}, \quad a_{\cdot j} = \sum_{k=1}^{M} a_{kj}, \quad \bar{a}_i = \bar{a}_{i\cdot} = \frac{1}{n}a_{i\cdot}$$

$$a_{\cdot\cdot} = \sum_{i,j=1}^{M} a_{ij}, \quad \bar{a} = \frac{1}{n^2}\sum_{i,j=1}^{M} a_{ij},$$

$|\boldsymbol{X}| = \langle\boldsymbol{X},\boldsymbol{X}\rangle^{\frac{1}{2}}$ is the Euclidean norm and $b_{ij}, \bar{b}_i, \bar{b}_j$ and $\bar{b}$ are defined similarly for $\boldsymbol{Y}$. It can be shown that this estimator for the distance correlation is unbiased with respect to $q$ and $p$.

**Global Coordination Level (GCL).** Consider an $N \times M$ matrix $\boldsymbol{X}$ representing the gene expression of $N$ genes from $M$ single cells, i.e., a matrix element $x_{i,j}$ represents the expression level of gene $I$ in cell $j$. To quantify the global coordination level, we first divide the genes into two random subsets, $S_1^k$ and $S_2^k$, each of them consists of $\frac{N}{2}$ genes. Then, we measure the dependency level $D^k = D_{bcDcorr}(X_1, X_2)$ where $X_1 = \{x_{i,j}\}_{j \in S_1^k}$ and $X_2 = \{x_{i,j}\}_{j \in S_2^k}$. In this study we use the bcdCorr described above, but essentially any high dimensional dependency measure can be used (e.g., Mutual Information). We repeat these steps $m$ times and define the GCL as the average of the dependency levels.

$$\mathrm{GCL(X)} \equiv \frac{1}{m}\sum_{k=1}^{m} D^k$$

In [24] it is shown that the GCL stabilizes for $m > 50$. Accordingly, in our analysis we choose $m > 50$.

**Gene Regulatory Network dynamics model.** The gene regulatory network (GRN) model [13] represents the dynamics of N interacting genes as a set of differential equations:

$$\frac{dx_i}{dt} = f_i \prod_{J_{i,j} \neq 0} g^s(x_j, \lambda_{i,j}, \theta_{i,j}, n_{i,j}) - k_i x_i$$

$g^s$ is the shifted Hill function,

$$g^s(y_j, \lambda_{i,j}, \theta_{i,j}, n_{i,j}) = \lambda_{i,j} + \left(1 - \lambda_{i,j}\right) * \left(\frac{1}{1 + (\frac{y_j}{\theta_{i,j},})^{n_{i,j}}}\right)$$

$\lambda_{i,j}$, is the maximum fold change in the production rate of node I that node j can cause.

Note that $\lambda_{i,j}$<1 if the edge inhibitory, and $\lambda_{i,j}$ >1 if the edge is activating. $\theta_{i,j}$, the threshold parameter of the Hill function and the hill coefficient $n_{i,j}$ were both set to 1. $k_i$ is the degradation rate of the regulator presented by node I, and $f_i$, is the production rate of the regulator presented by node I was chosen to be 1. The connectivity matrix non-zero elements, J, represents the network topology, were chosen to be either random network, or a more realistic gene regulatory network (SI Fig. 4a), with an average network density C. The regulatory interactions, $\lambda_{i,j}$, that was assigned to the non-zero connectivity matrix $J$, was chosen according to the symmetric distribution: $\lambda_{i,j}, \sim e^{-randn(0,W)}$, where an equal number of repressive and activating interactions were drawn, and W, governs the weights mean and variability.

To simulate cohorts with varying level of interaction intensity, we either change $W$, which controls of the mean and variability of the network weights, or we change the network density $C$. For a specific GRN model, defined by $\lambda_{i,j}$, the expression profile of a ‘single cell’ in a ‘cohort’ is generated by integrating the GRN differential equations with adding stochasticity to different parameters: 1) The initial gene expression levels $x(t = 0)$ were randomly drown from a uniform distribution $\mathcal{U}(0,1)$. 2) We added stochasticity to the weights by adding a random variable selected from a normal distribution with STD of $\sigma_n$ to a fraction $P_{\text{stochastisity}}$ of the interactions such that $\lambda_{i,j\epsilon P}, \sim e^{-randn(0,W)+stochastisity(0,\sigma_n)}$. We evaluate the steady state using the ode45 MATLAB function.

**Diversity of expressed genes.** We estimate the diversity of expressed genes in each cell gene expression profile and define it as the effective number of expressed genes using Shannon diversity ($N_{\text{eff}}$), which quantifies both the number of expressed genes and their dominance in the cellular gene expression profile. $N_{\text{eff}} = e^{-\sum_{i=o}^{S} p_i \ln p_i}$, where $p_i$ is the fraction of counts of gene i, from $S$ expressed genes. We also tested alternative definitions: (1) The number of observed genes with more than one count. And (2) the number of most expressed genes that sum up to 90% of the total transcription.

***GCL*-$N_{\text{eff}}$ tradeoff in gene regulatory network dynamics.** In order to test the effects of network interaction intensity (e.g. network density or weights) on the effective number of genes in the system's stable states, we simulate *M=30* cohorts of cells, producing *N=100* stable states in each cohort, originating from varying interaction intensity GRNs, defined by different pairs of *C* and *W*. For every GRN, we initiate with a size of 300 genes, and set the number of non-zero elements $\lambda_{i,j}$, setting the network density *C*, and assigning its average weight, while preserving its network topology cross samples. When increasing the network density, we are adding links to the same initial network. After 'ode45' run we calculate both GCL and $N_{\text{eff}}$ to cell cohorts which satisfies the gene expression stability test in all cells.

**Cell filtering procedure.** Gene expression is represented as counts, and after the first step of the filtering process, we normalize the counts in each cell according to the sum of all counts, such that for each cell, $\mathrm{x}_{\mathrm{I,j}} = \frac{\mathrm{counts}_{\mathrm{I,j}}}{\sum_j \mathrm{counts}_{\mathrm{I,j}}}$. We applied the same following process across all cell types and all the datasets: (a) Remove cells which the expression profile of their top ten most highly expressed genes was an outlier. We calculated per cell, the fraction of the counts sum of the top 10 most highly expressed genes from the total counts in the cell, then, we removed cells exhibiting fractions exceeding two STDs. (b) Filter cells for which the number of expressed genes ($x_{i,j} > 0$) is larger or smaller than two standard deviations from the mean number of expressed genes. (c) Filter cells for which the effective number of expressed genes ($N_{\text{eff}}$, as defined in *Methods*) is larger or smaller than two standard deviations from the mean $N_{\text{eff}}$. (d) *Remove small clusters:* We apply Louvain algorithm for cells clustering detection on the Spearman distance network of cells, followed by Silhouette analysis which yields a score $s_i, (i = 1, \dots, N)$, for each cell and sum the scores $S_{Louvain} = \sum_{i=1}^{N} s_i$. In case we find cluster with $S_{Louvain} > 0.25$ we only keep the largest cluster and repeat this process again to validate that there are no further clusters. We also apply $t$-SNE dimensional reduction visualization as a complementary analysis. We used these techniques to detect the presence of clusters in the data. If clusters are detected, we keep the largest cluster, detected using the Louvain approach detailed above. We Calculate the Spearman distance between each cell and the mean cell and remove cells with a distance larger than two standard deviations above the average distance. (f) Calculate the Spearman distance between all pairs of cells and remove cells with distance smaller than four standard deviations below the average distance. (g) In order to further remove small clusters that were revealed only after the removal of outliers, we repeat the clustering filtering approach from step (d) and finally, apply a k-means approach over the PC1-PC2 plane from PCA. See SI Fig. 13 for a step-by-step example of the pre-process procedures.

**Standard GCL and $N_{\mathrm{eff}}$ analysis for all the data sets.** Each dataset includes samples coming from different tissues, locations, and sometimes from different ages or sex. We considered a sample as a cohort of cells that came from a specific tissue, cell type and a specific donor/mice/worm/fruit fly and its unique age and sex. For cell type annotation, we used the given annotations of the analyzed dataset. In order to avoid possible counts effects on the calculation of the GCL and $N_{\mathrm{eff}}$, we down sampled each dataset separately, as described in "Down-sampling process" in methods. We used raw counts data in all datasets not using "within sample" normalization in order to avoid biases in GCL calculation [45,46]. For each dataset, the GCL is calculated per sample, following these steps: (a) filter cells as described in the "cell filtering procedure" methods section, focusing on cell cohorts with more than 40 cells. (b) Sort the genes according to their mean gene expression level, focusing on the top 1,000 highly expressed genes (in main text figures). (c) We perform bootstrap analysis by selecting 100 random subsets of genes, with repeats, each subset consists 50% of the number of cells in the cohort, and calculate the GCL for each subset. Finally, the gene-to-gene coordination of each cohort is represented by a distribution of 100 GCL values. After characterizing all cell cohorts, we (a) focus on cell cohorts where the mean number of counts in these cohorts, did not exceed two STDs from the mean number of counts in all cohorts. (b) We removed cell cohorts which their GCL values were lower than the 2 STDs GCL of the shuffled cohort scRNA-seq data.

**Pathway analysis.** *Pre-processing*: We analyzed 330 pathways as annotated by KEGG [29]. The R programming language and the biomaRt package [47,48] are used to convert the gene identifiers to those used in the different analyzed datasets. We applied the pathway analysis the top ten most highly coordinated pathways in human Bone marrow hematopoietic stem cells, and eye photo-receptor cells, and in mice Aorta fibroblasts of cardiac tissue and bladder urothelial cells. We filter genes which are expressed ($x_{i,j} > 2$) in less than 20% of the cells of the cohort. Finally, we only analyze pathways that consist at least 20 genes.

*Surrogate analysis*: We calculate the GCL of each pathway and compare it to the GCL values of 20 corresponding 'surrogate' gene-sets, yielding a Z-score for each pathway. The surrogate gene-set are selected as follows: Consider a set of $n$ genes $G = \{g_1, \ldots, g_n\}$ that are annotated as a particular KEGG pathway. First, all $N$ genes are ranked according to their mean expression, where $k_i$ ($i = 1, \ldots, N$) represents the rank of gene $i$. Next, we choose a same-size surrogate gene-set $\tilde{G} = \{\tilde{g}_1, \ldots, \tilde{g}_n\}$ such that for each gene $g_j$ from the original pathway ($g_j \in G$) we randomly choose a gene $\tilde{g}_j$ with the constraint that its expression rank $k_{\tilde{g}_j}$ is similar to the rank $k_{g_j}$ of original gene $(k_{g_j} - 5) \geq k_{\tilde{g}_j} \geq (k_{g_j} + 5)$ and $\tilde{g}_j \neq g_j$. The resulted surrogate gene-set $\tilde{G}$ preserves the size of the original pathway $G$ and to mimic its expression profile but does not represent any known KEGG pathway. Finally, $\text{GCL}_{surrogate}$ values are calculated for 20 surrogate gene-set $\tilde{G}^{(\nu)}$ $(\nu = 1, \dots, 20)$. See schematic demonstration in SI Fig. 2. We then calculate a Z-score, defined as $z \equiv \frac{(\text{GCL}_{pathway} - mean(\text{GCL}_{surrogate}))}{std(\text{GCL}_{surrogate})}$.

**Down-sampling process.** Different datasets have used different scRNA-seq techniques, and within each dataset, different pipelines and different constraints are associated with different tissues and cell types, resulting in different scRNA-seq sequencing depth for different cell types and tissues. These differences may affect the analysis of both GCL and measuring the effective number of expressed genes in each cell types. One of the ways in which we tried to avoid counts effects on the calculation of the GCL and $N_{\text{eff}}$, while comparing cell cohorts from different cell types, and from different scRNA-seq sequencing pipelines, is to down sampled cells in each dataset to a constant level. The down sampling procedure was the same for all datasets. Different down-sampling desired number of counts was chosen per dataset, to be the lowest number of counts that would allow us to analyze at least 95% of the cells. The down-sampling process was done in a way that preserved the distribution of counts across genes in each cell, and at the end of the down-sampling process, all cells in the dataset had the exact number of counts **n**. The fundamental concept of the down-sampling process involves partitioning the interval [0, 1] into **K** bins, wherein the size of each bin corresponds proportionally to the allocated probability mass of the number of counts distribution in each cell. Subsequently, **n** values are sampled from a uniform distribution within the range of [0, 1]. The outcome is determined by identifying the bins into which these generated values fall. The bins that encompass these values are then selected as the results of the process. This method ensures that the probability distribution is accurately reflected through the sampling procedure, providing a reliable means of generating representative outcomes.

**Statistical test for the $GCL$ -$N_{\text{eff}}$ tradeoff pattern.** To assess the statistical significance of the observed patterns of GCL versus $N_{\text{eff}}$ (Fig. 2-3), we conducted two independent tests. Firstly, we examined the overall trend by calculating the Pearson correlation coefficient and its associated P-value to determine the presence of a negative relationship between GCL and $N_{\text{eff}}$. Secondly, we evaluated the constrained complexity of the GCL – $N_{\text{eff}}$ pairs by establishing an empirical upper boundary curve and determining its statistical significance compared to randomized realizations. To define the empirical upper boundary curve, we employed a curve of the form $F(x) = \frac{\alpha}{1+e^{-\beta(x+\gamma)}} + \varepsilon$ where $\alpha, \varepsilon$, β, and γ are fitting parameters. $\alpha = \max(GCL)$, $\varepsilon = \min(GCL)$. The curve β, and γ are fitting parameters that were selected to ensure that at least 95% of the points ($GCL < N_J$, where j represents all GCL – $N_{\text{eff}}$ pairs) lie below it. Moreover, we aimed to achieve this while maintaining a minimal ratio, denoted as r, between the number of data points and the number of possible GCL – $N_{\text{eff}}$ combinations below the curve. The possible GCL-$N_{\text{eff}}$ combinations encompass all feasible pairs of GCL and $N_{\text{eff}}$, independently selected from the observed values. To determine the statistical significance, we utilized the test statistic r. We repeated the process for 100 Monte Carlo realizations of shuffled data, where the $N_{\text{eff}}$ values were randomly assigned to the GCL values. The Z-Score was then calculated between the shuffled realizations r values distribution and the r that was observed in the actual data.

**BMDM growth, stimulation and sequencing.** BMDMs were differentiated from mouse bone marrow cells in macrophage growth media (MGM): 70% complete RPMI (RPMI-1640 with 10% heat-inactivated fetal bovine serum, 1 mM sodium pyruvate, 200 U/ml pen/strep, 2 mM L-glutamine and 10 mM HEPES) and 30% L929-cell conditioned media as a source of the macrophage growth factor, M-CSF. Bone marrow was extracted from the femurs and tibias of C57Bl/6J mice by crushing in PBS, followed by ACK (ammonium-chloride-potassium) lysis of red blood cells and passage through a 70-μm cell strainer. Bone marrow from one mouse was split between 5 non-culture treated 15 cm dishes in 25 mL of MGM (Day 0). On day 3, cells were supplemented with 10 mL of MGM. After day 6, adherent cells were lifted with cold phosphate-buffered saline (PBS) containing 5 mM EDTA, washed and 1 million cells/ml were plated in 10 cm non-culture treated dishes in 10 mls of MGM. Dishes were stimulated with 10 ng/ml LPS (Sigma), IL-4 (R&D) or left unstimulated for 2 hours. Cells were then lifted with cold phosphate-buffered saline (PBS) containing 5 mM EDTA and sorted into 96 well dishes containing 10ul of TCL. Libraries were constructed as previously described [49]. Libraries were sequenced on a Nextseq 500 and fastq files were processed using Kallisto [50] to obtain estimated counts used in this analysis.

## Data availability

All data sets analyzed in this study are publicly available as described in Extended Data Table 1. The MATLAB code for computing the GCL, and any other analysis, will be shared by the authors upon request. Macrophages stimulation data will be available upon request.

## Author contributions

R.M., O.L. and Y.Y.L. conceived and designed the project. O.L. performed scRNA-seq data analysis, the simulations and bio-informatics analysis of pathways. S.D.P applied the Macrophages experiment and bio-informatics analysis of stimulation data. S.T. Applied frustration analysis and simulations. All authors analyzed the results. R.M., O.L. and Y.Y.L. wrote the manuscript with contributions from all authors.

## Competing interests

The authors declare no competing interests.

## Additional information

SI figures 1-13 and SI tables 1-2

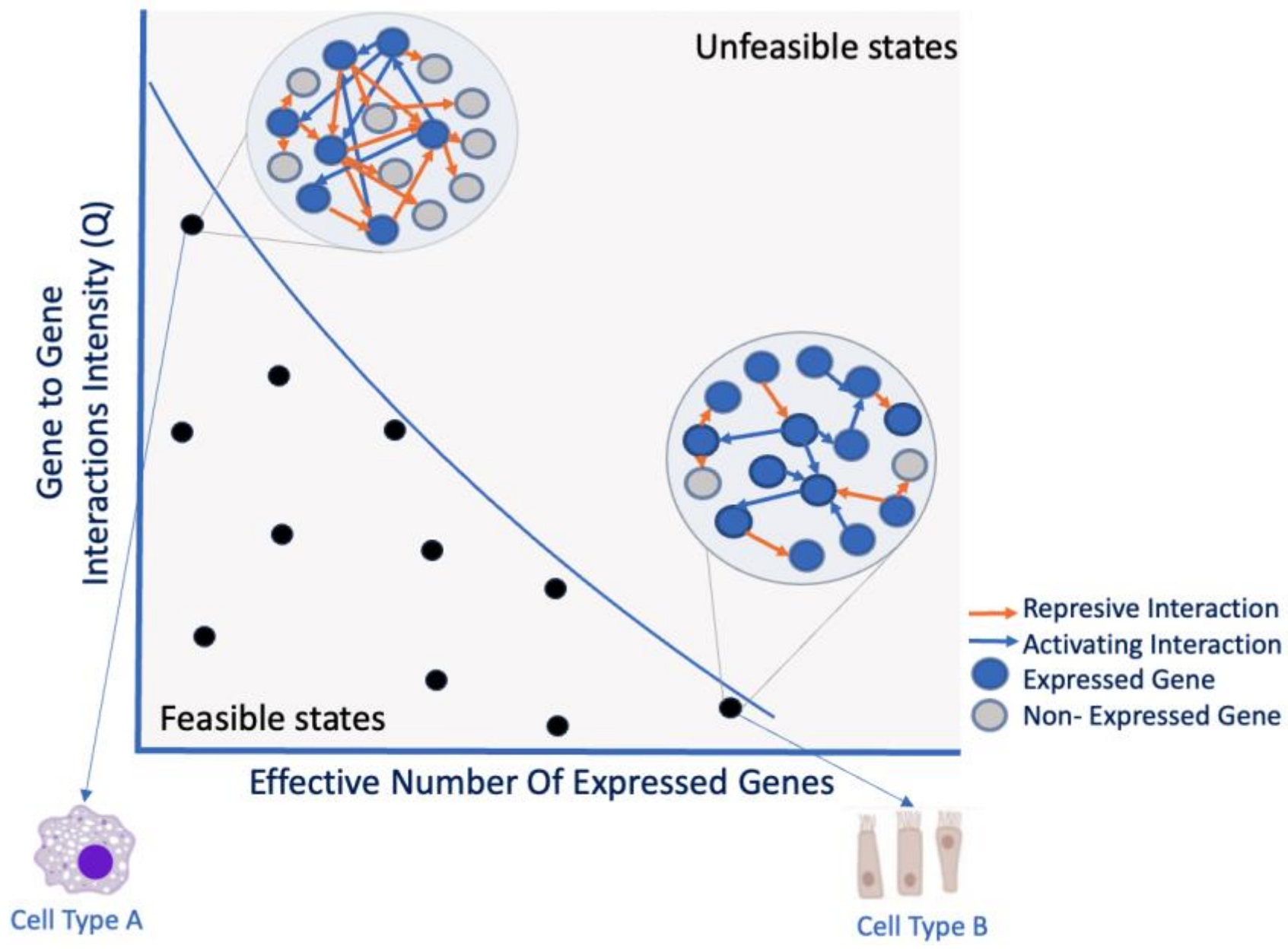

**Fig. 1 | The intensity of gene regulatory interactions limit the diversity of gene expression.** An illustration of our hypothesis that high interaction intensity limit the diversity of gene expression. We hypothesize that there is an upper limit to the overall complexity, e.g., multiplication of the number of genes and their interaction intensity of the gene regulatory network (GRN). Each dot represents a different cell type, with two characteristics: effective number of expressed genes ($N_{\text{eff}}$), and its GRN interactions intensity, $Q = W * C$, where $C$ represents the GRN interaction density and W its average interaction weight. According to the hypothesis, it would be improbable to observe cell types with both a large number of genes and a high level of gene-to-gene interaction intensity. While the effective number of expressed genes is, in principle, an observable measure, the networks of gene-to-gene interaction intensity are generally unknown. Dots in the $Q - N_{\text{eff}}$ plane that are below the critical feasibility curve represent feasible gene regulatory networks. In this study, we analyze scRNA-seq samples of different cell types, collected from different tissues and different organisms. These samples represent diverse GRNs and gene expression profiles. Our goal is to better understand the limitations of complexity, a product of the diversity with the interaction intensity, in natural gene regulatory networks. In this illustration, we demonstrate two different cell types, Cell type A and B, which have large, and low effective number of expressed genes. Cell type A expresses many genes, and its underlying gene-to-gene interactions intensity are low, while Cell type B which express a small number of genes have a network with high interaction intensity. We suggest that the GRN complexity is bounded and that the diversity of expressed genes is thus limited.

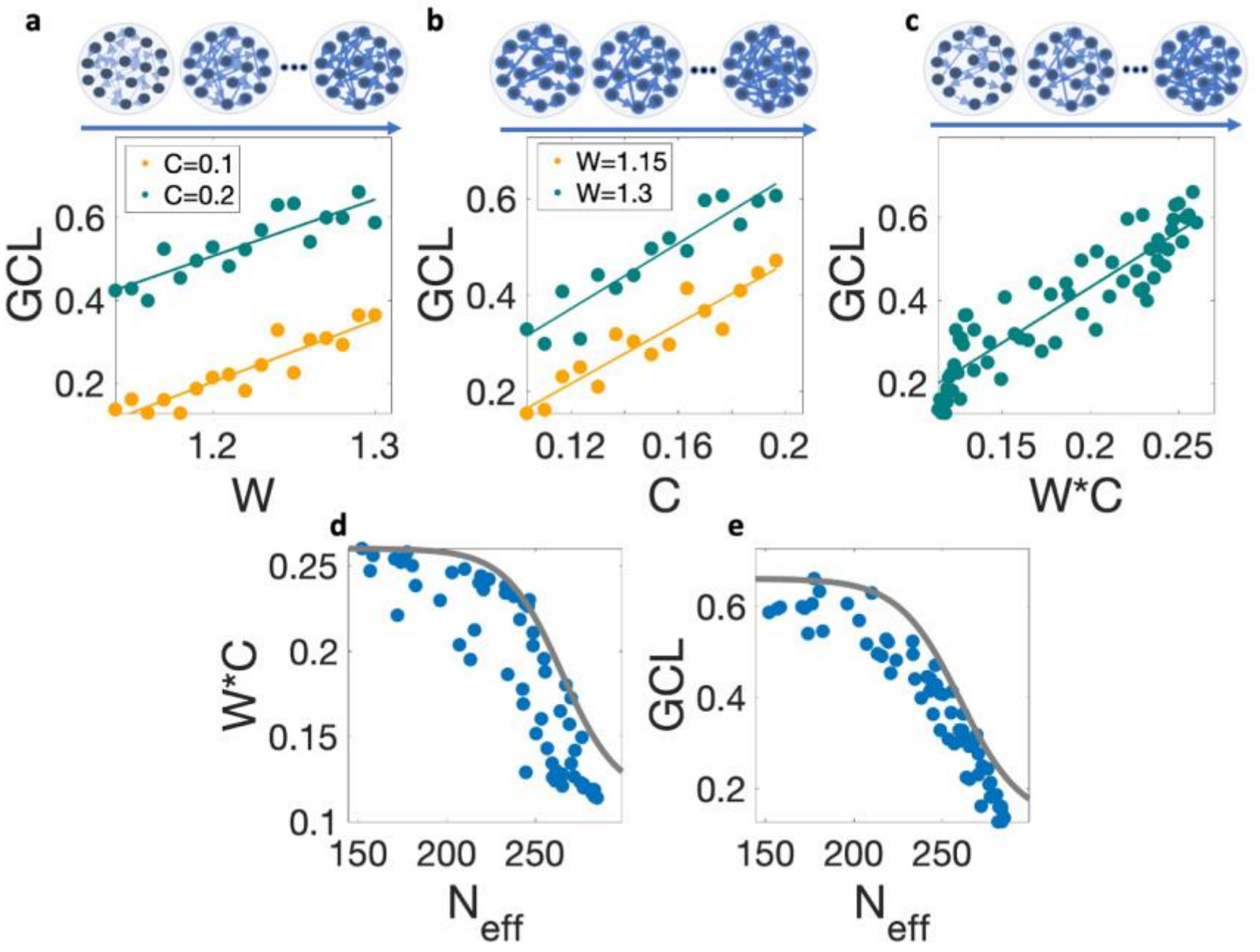

**Fig. 2 | GCL analysis of simulated cell types reveals the predetermined interactions intensity of the underlying GRN dynamics and diversity - GRN intensity tradeoff.** We demonstrate the calculation of GCL using Co-regulation gene regulatory network (GRN) dynamics model, with a random network topology, and different underlying gene-to-gene interaction intensities (e.g., different weights or network density). The schematic networks in **a,b,c** demonstrates different forms of increased network intensities. **a**, Increasing the network intensity by increased weights. We generate $N = 30$ samples with average weights $W = 1{:}\,0.01{:}\,1.3$, network of 300 genes, and an average network density $C = 0.1$ and additional $N = 30$ samples of $C = 0.2$. For each sample with sets of $C, W$ we generated *M=100* stable states and calculate the GCL and $N_{eff}$. We find Pearson correlation of 0.91 ($P_{val}$<0.001) for $C = 0.1$ and of 0.97 ($P_{val}$<0.001) for $C = 0.2$ between the underlying GRN weights and the GCL. **b.** Increasing the network intensity by increased network density. We generate $N = 30$ samples with average density varies from 0.1:0.003:0.2, network size of 300 genes, and an average weight $W = 1.15$ and $W = 1.3$. For each sample we generated M=100 stable states, we find a strong Pearson correlation of 0.98 and 0.97 ($P_{val}$<0.001) between the underlying GRN density and the GCL. In **c**, we integrate the sets of parameters from **a,b** changing both *W* and *C*, and present their multiplication values and find a strong Pearson correlation of 0.98 ($P_{val} < 0.01$). In (**d**)**,** we calculate the interactions intensity of the underlying GRNs which is represented by the multiplication mean value of W and the density C, and calculate the effective number of genes ($N_{Eff}$), while increasing the interactions intensity as in **c,** and find a strong negative Pearson coefficient of -0.93 ($P_{val} < 0.01$). In **e**, we demonstrate the GCL ability to capture this effect from the systems stable states, which may be measured in real scRNA-seq data. In both **d,e**, The gray line corresponds to the empirical upper limit boundary curve of the data points, where the Z-score is calculated as 7.84 and 9.02 in **d,e**, respectively, using the reshuffling test (described in the Methods section). This calculation is based on 200 Monte Carlo simulations.

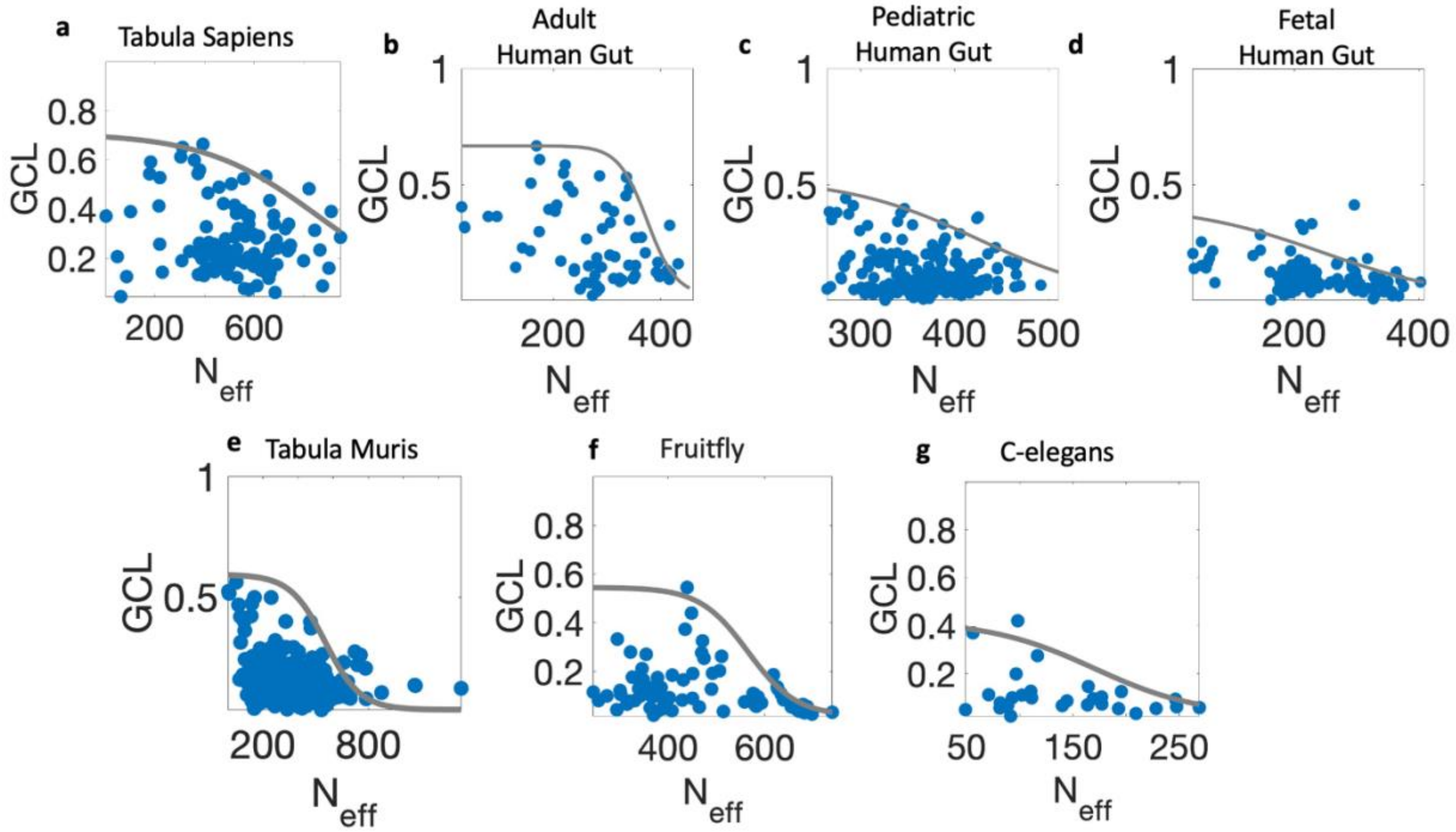

**Fig. 3 | $GCL$-$N_{eff}$ trade-off observed in real scRNA-seq data.** Cross tissues and cell types scRNA-seq data analysis of GCL and $N_{\mathrm{eff}}$ of different organisms. The dots represent different cohorts of single cells genes expression data from different cell types, collected from various organisms. Each cohort of cells obtained from a specific cell type and had more than 40 cells. **a,** Samples collected from human donors [31], 114 cell cohorts were analyzed (*See methods*), each with more than 40 cells. The Pearson correlation rank is −0.2 ($P_{val}$ < 0.01). The gray line corresponds to the empirical upper limit boundary curve of the data points, where the Z-score is calculated as 2.4 using the reshuffling test (described in the Methods section). This calculation is based on 200 Monte Carlo simulations. **b,c,d,** Human gut cross sectional samples [33] from: healthy adults, pediatric and fetal donors, with 66, 129 and 214 cell cohorts respectively. The Pearson correlation rank is −0.34 ($P_{val}$ < 0.001), −0.27 ($P_{val}$ < 0.001), −0.18 ($P_{val}$ < 0.005) and reshuffling test Z-score of 2.54, 2.65 and 3.05. **e,** Mice cross tissues samples [32], 218 cell cohorts were analyzed. The Pearson correlation rank is −0.32 ($P_{val}$ < 0.001) and Z-score=2.81. In **f.,** Fruit-fly cross tissue samples [34] were analyzed, 74 cell cohorts, Pearson correlation rank is −0.23 ($P_{val}$ < 0.05) and Z-score=4.96. **g.,** C-elegans cross tissue samples [35] were analyzed, 30 cell cohorts, Pearson correlation rank is −0.29 ($P_{val}$ < 0. 1) and Z-score=2.76. Overall, all of the analyzed exhibiting significant GCL-$N_{\mathrm{eff}}$ tradeoff pattern with similar behavior to the simulations.

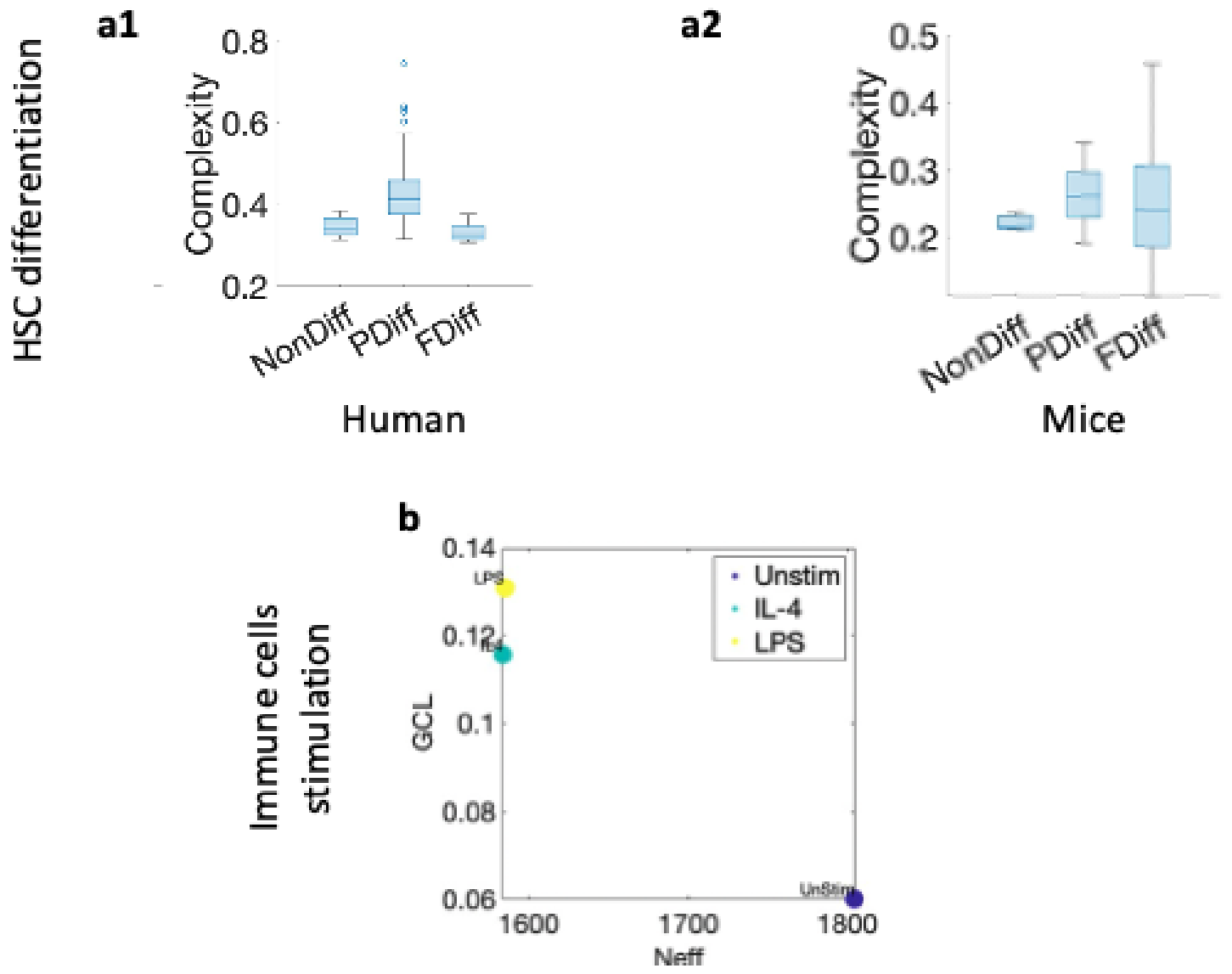

**Fig. 4 | GRN intensity - diversity tradeoff patterns upon differentiation and stimulation. a.** Complexity comparison of HSCs in different differentiation stages from marrow and peripheral blood from two human and mice were analyzed. We divided the cell cohorts to three groups according to their differentiation stages: Non-Differentiated (NonDiff stem cells), Partially differentiated (PDiff, transition state cell types) and fully differentiated (FDiff). In human we analyze 4, 51 and 22 cohorts and in mice 3,7 and 12 cohorts of NonDiff, PDiff and FDiff respectively. The complexity was calculated by a multiplication of the normalized GCL and $N_{eff}$ values. **b.** We compare $N_{eff}$ and GCL of Macrophages LPS and IL4 stimulated cell cohorts and unstimulated macrophages. While expecting an increase in the diversity of expressed genes due to the stimulation, we observe a clear decrease in the diversity of expressed genes, and a pronounced tradeoff between the GCL and $N_{\text{eff}}$.

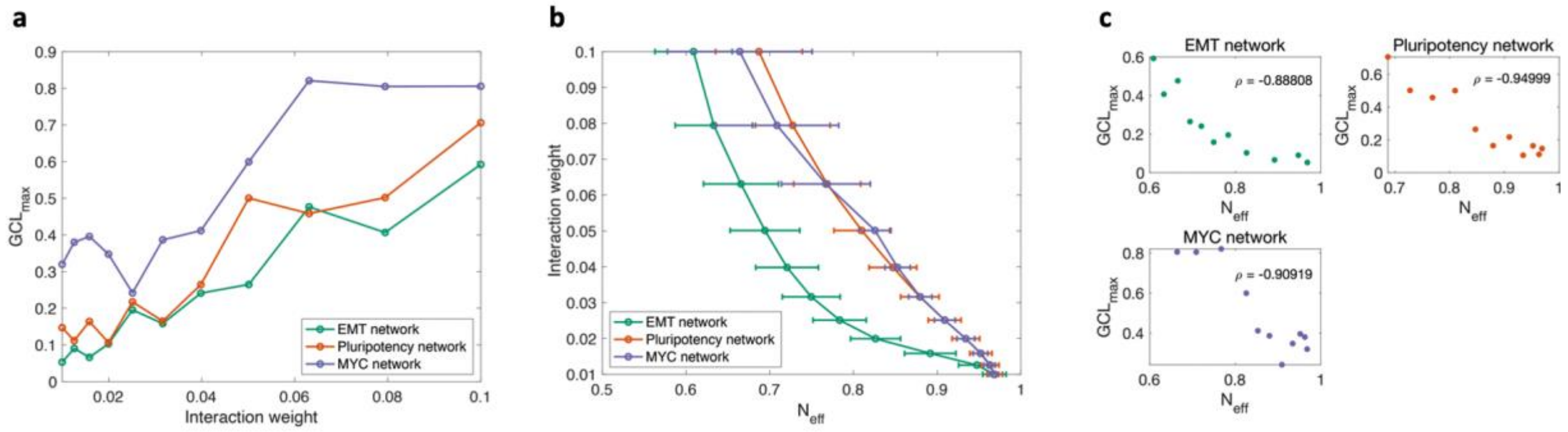

**SI Fig. 1 | GCL-$N_{\text{eff}}$ tradeoff is also seen in the case of biological networks.** Analyzing realistic networks taken from the literature and involved in different cellular processes: (1) the EMT (epithelial-mesenchymal transition) network which governs cell-fate choice between epithelial and mesenchymal cell fates [27], (2) the pluripotency network which governs human embryonic stem cell differentiation [37], and (3) a regulatory network involved in regulating the choice between different MYC gene-associated breast cancer cell phenotypes, inferred from gene expression profiles of breast cancer patients [28,13]. (A) GCL increases with the network interaction intensity in the case of three biological gene regulatory networks. Here, we keep the network density (i.e., the number of edges in the gene network) fixed and vary the weight (or strength) of the network interactions to change the network interaction intensity. See Supplementary Methods for details of the simulation setup. (B) $N_{\text{eff}}$ calculated for the setup in (A). Error bars indicate the standard deviation. (C) The GCL is negatively correlated with $N_{\text{eff}}$ in the case of all three biological networks. The GCL and $N_{\text{eff}}$ values calculated in panels A and B are re-plotted in panel C; each dot corresponds to the network behavior for a fixed interaction intensity. The Pearson correlation coefficient between GCL and $N_{\text{eff}}$ is shown in each subplot.

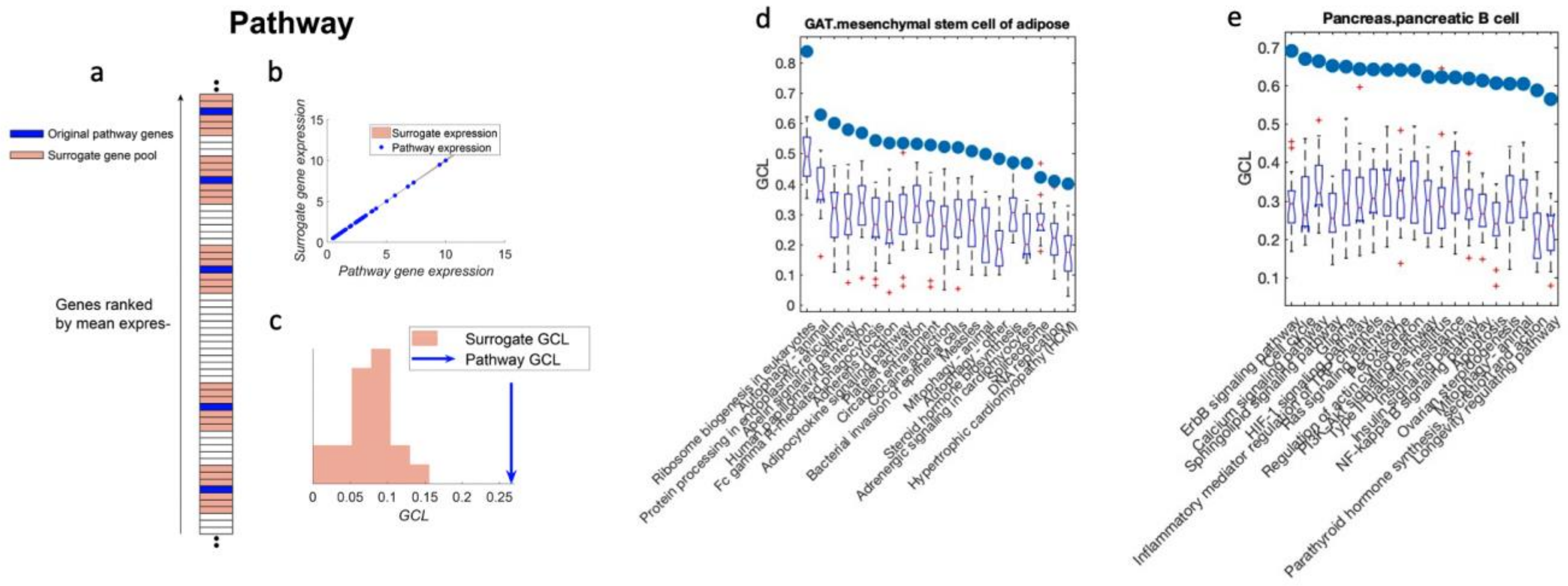

**SI Fig. 2 | GCL captures effect of gene-to-gene dependency and functional relations in real scRNA-seq data. a,** Visualization of the surrogate preparing process. Each pathway is composed of several genes with different mean expression values. We sort the genes by their mean expression values and for each gene belonging to the pathway (blue) we identify a subset of genes with similar expression values (pink). A surrogate expression profile is generated by selecting a random gene from each subset. This allows us to generate many surrogate pathways with similar expression values. **b,** the expression levels of the ‘T cell receptor signaling pathway’ are shown versus the mean values of the corresponding surrogate gene sets (n=20). The red area represents the range of expression values of the surrogate gene sets, demonstrating their similarity to the real pathway. **c,** GCL values of the real pathway (blue arrow) and the surrogate pathways (red histogram). Even though the expression profiles are very similar, the GCL of the real pathway is significantly higher than the surrogate pathway ($P < 0.05$). This suggests that higher coordination is associated with real biological function. **d-e,** GCL values selected pathways are compared with GCL values of compatible surrogate gene-sets, yielding a Z-score, by which the pathways are sorted in the figure. We present the top 20 most highly coordinated pathways in GAT mesenchymal stem cells of adipose, and pancreas pancreatic B cells from Mice Tabula Muris dataset [32]. Note that both cell types exhibit pronounced fraction of significant number of elevated GCL (76 and 49) compared with the random case (24 out of 150 relevant pathways, which had sufficient gene expression profile).

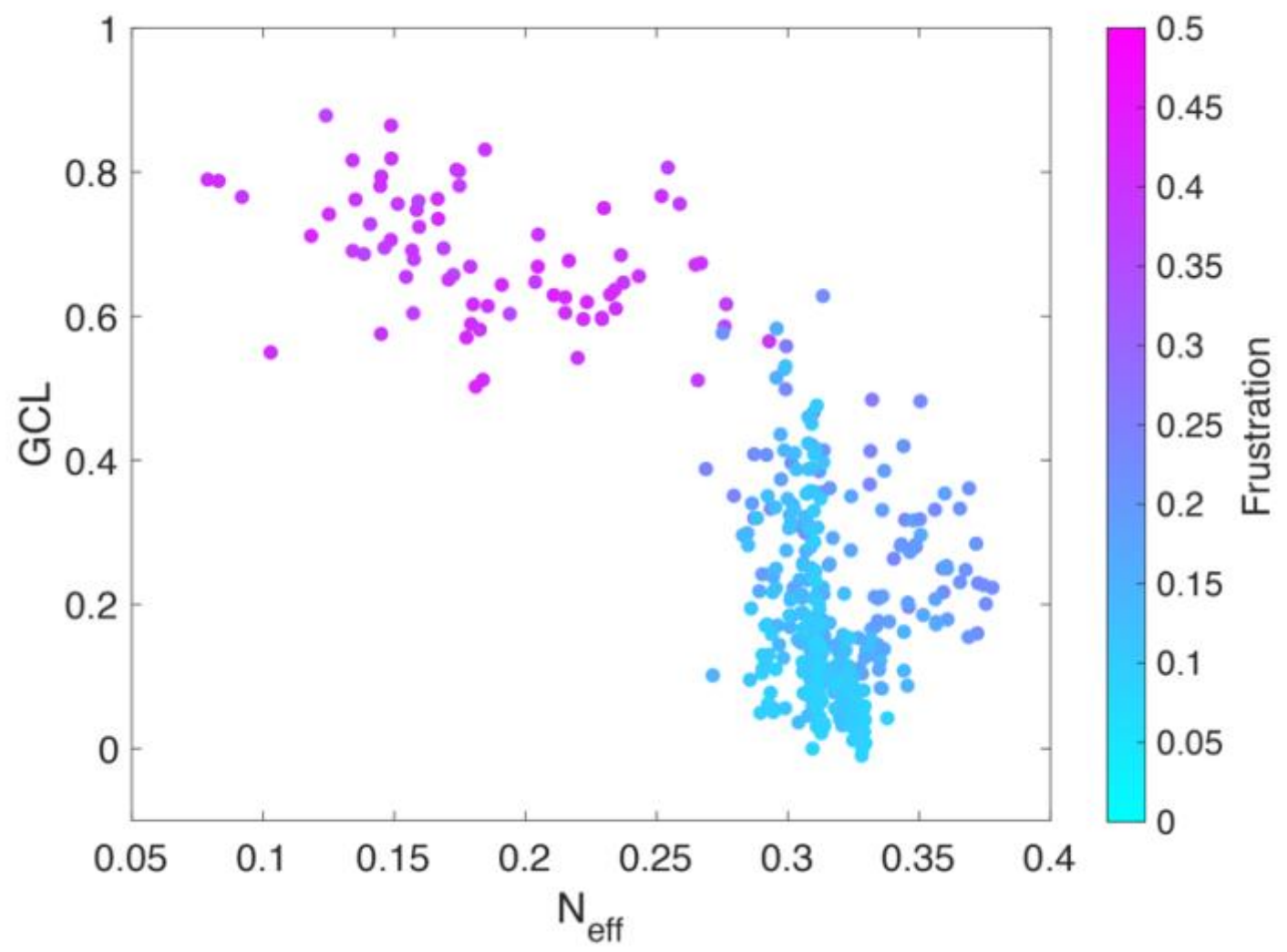

**SI Fig. 3 |** The GCL-$N_{\text{eff}}$ tradeoff is also seen in the case of an ensemble of gene regulatory networks exhibiting the minimal frustration property, a characteristic of biological regulatory networks seen across contexts [13]. The minimally frustrated networks analyzed here were obtained by starting with a collection of random networks and adding edges to the networks under selection for the minimal frustration property (see Supplementary Methods for details of the procedure used to obtain the networks). The various networks analyzed here have the same number of nodes but differ in the number of edges. Each dot shown in this figure corresponds to one collection of steady states obtained from simulations of a network in the ensemble. The color of the dots indicates the frustration of the steady states in that collection of states (see color bar). For this collection of minimally frustrated (but otherwise random) gene regulatory networks, we recapitulate the negative correlation between GCL and $N_{\text{eff}}$. The figure also shows that states with high GCL and low $N_{\text{eff}}$ are more frustrated. Such states have previously been shown to be less stable [13]). Note that the networks analyzed here were obtained from a collection of random networks without any explicit consideration of the GCL or $N_{\text{eff}}$ values: the GCL-$N_{\text{eff}}$ tradeoff and the association of GCL and $N_{\text{eff}}$ values with frustration of the network states seen here emerges naturally.

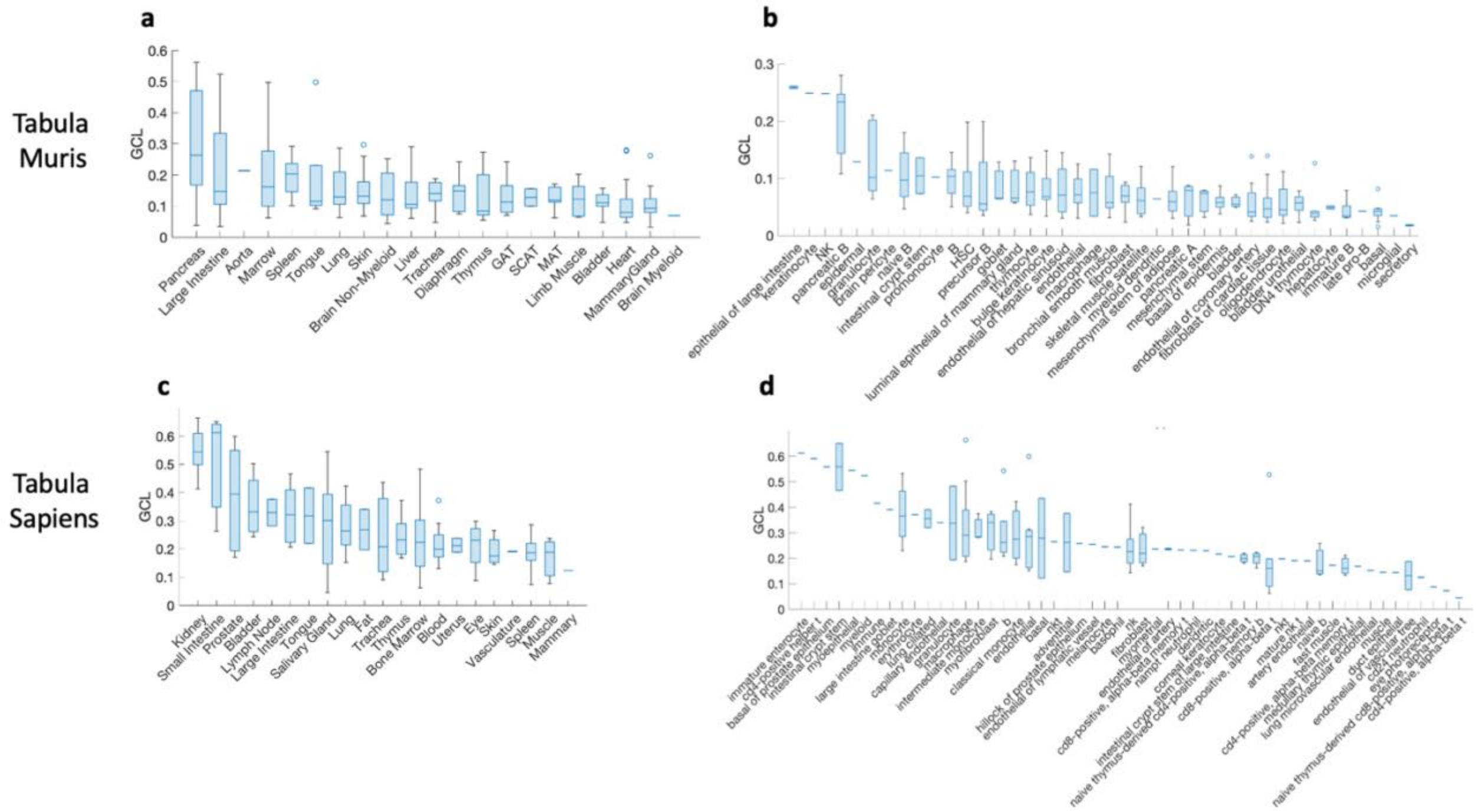

**SI Fig. 4 | GCL of different cell types in diverse tissues of Humans and Mice.** GCL calculation of real single-cell RNA seq data of cohorts from a wide range of tissues and cell types obtained from human and mice. **a,b**, Tabula Muris mice data was analyzed, GCL was calculated over 218 cohorts of cells, each from a given cell type. In **a,** we introduce the GCL values of cell types from different tissues. The GCL values that are presented are the mean and STD of different cell types from a given. In **b,** the GCL values in each cell type variates from different tissues and different mice. **c,d,** Tabula Sapiens human data was analyzed, GCL was calculated over 114 cohorts of cells, sorted by tissues (**c**) and by cell types (**d**)**.**

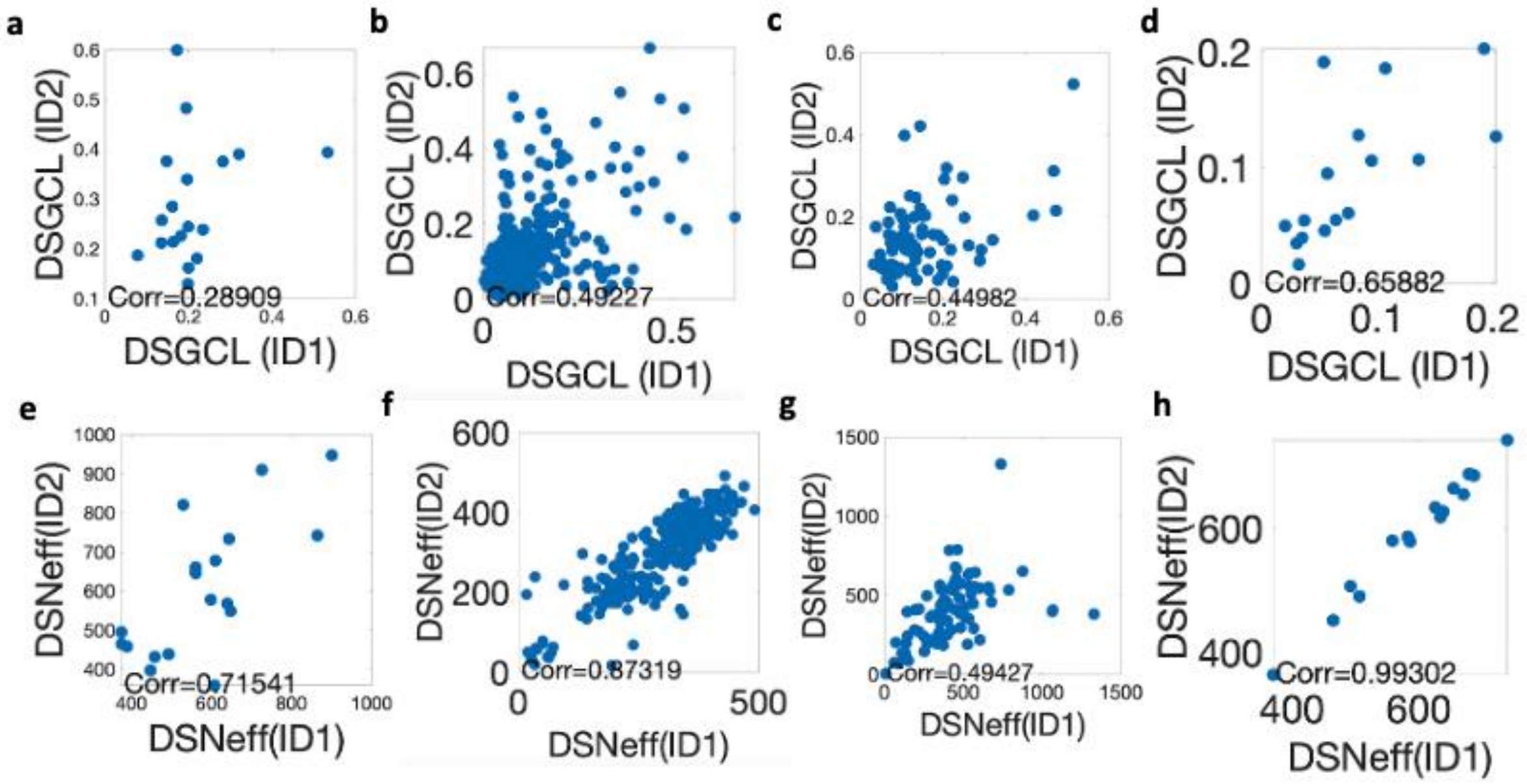

**SI Fig. 5 | Consistency of GCL and $N_{\text{eff}}$ cross cell types in different datasets.** Testing the consistency of GCL and $N_{\text{eff}}$ values of different cell types across samples obtained from different mice/human/fruit fly. For each cell type, we compare the GCL values of different samples from the same age and the same tissue **a,e,** Samples collected from human donors. **b,f,** Human gut . **c,g,** Mice Tabula Muris. **d,h,** Fruit-fly cross tissue samples were analyzed.

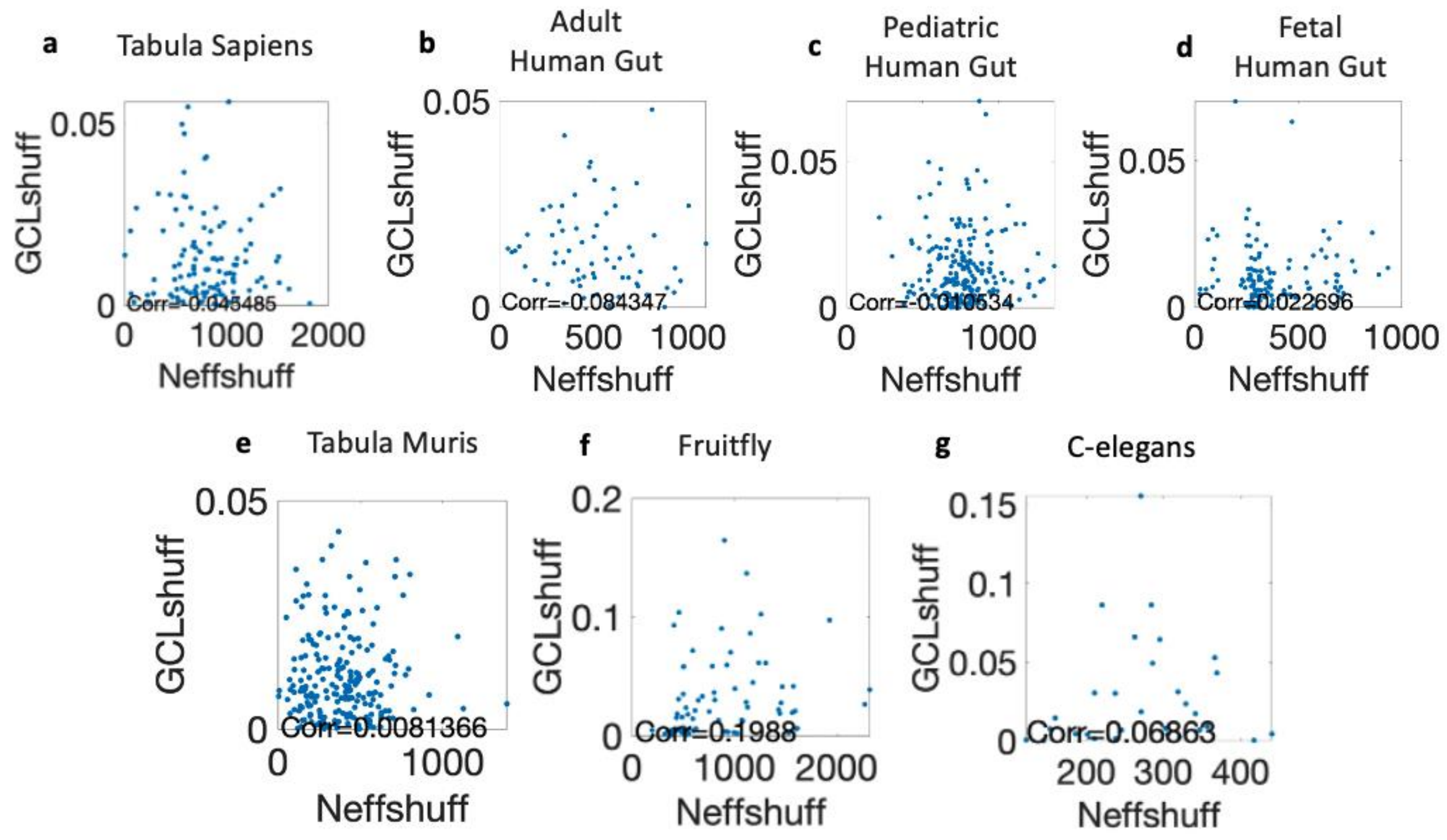

**SI Fig. 6 | Shuffled GCL and $N_{\text{eff}}$ cross cell types in different datasets.** Cross tissues and cell types shuffled scRNA-seq data analysis of GCL and $N_{\text{eff}}$ of different organisms. The dots represent different cell type cohorts, each cohort was shuffled separately, where the gene expression distribution of each gene was preserved. As in Fig. 2, each cohort of cells obtained from a specific cell type and had more than 40 cells. **a,** Samples collected from human donors, the Pearson correlation rank is 0.008 ($P_{val}$ > 0.05). **b,c,d,** Human gut cross sectional samples from: healthy adults, pediatric and fetal donors. **e,** Mice cross tissues samples. In **f.,** Fruit-fly cross tissue samples were analyzed. **g.,** C-elegans cross tissue samples were analyzed. Overall, all of the analyzed shuffled data exhibit no-significant GCL-$N_{\text{eff}}$ tradeoff pattern where the correlation is weak positive in all datasets, all with non-significant $P_{val}$> 0.05.

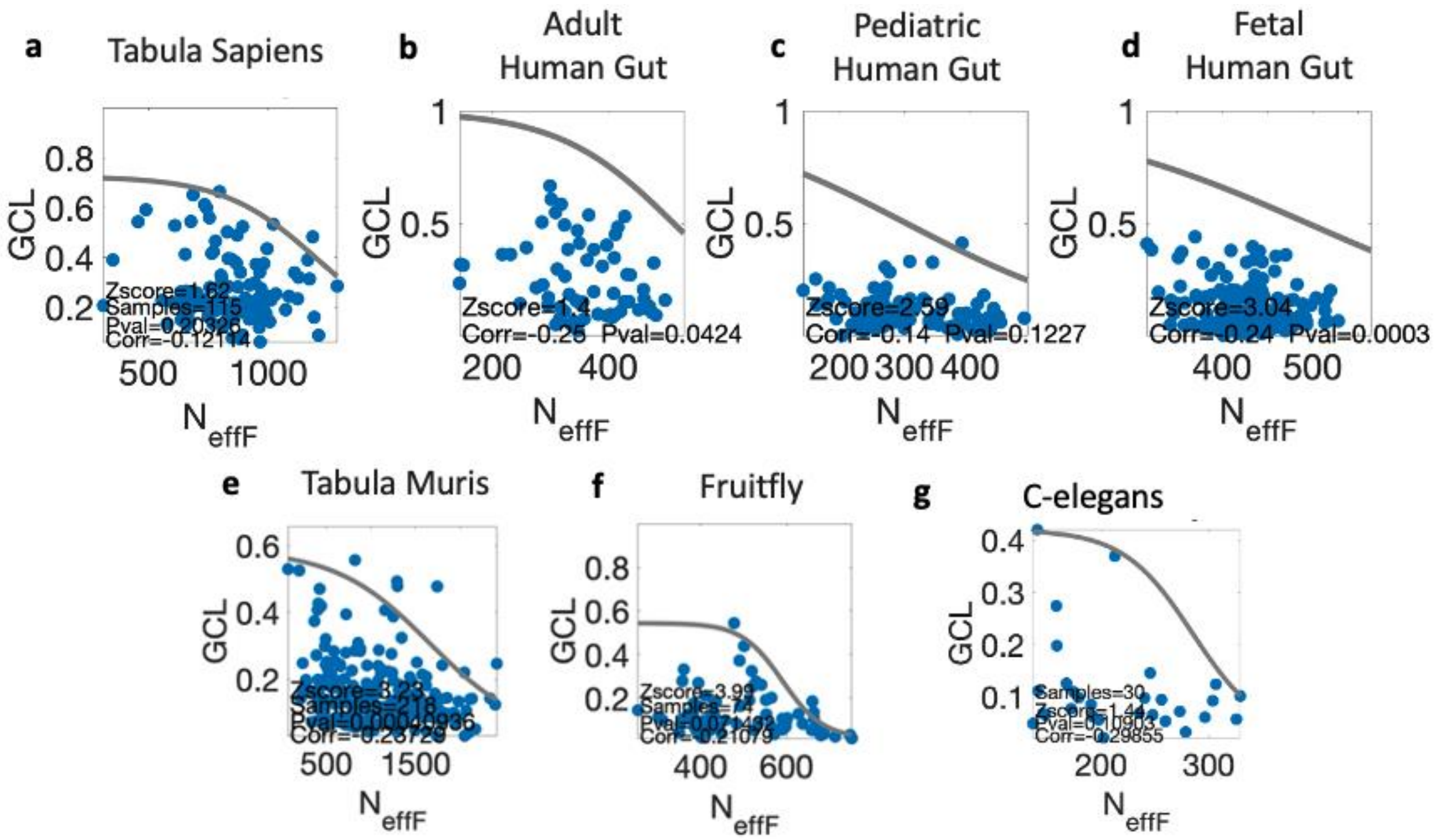

**SI Fig. 7 | $GCL$-$N_{\mathbf{effF}}$ trade-off observed in real scRNA-seq data.** Here we present the robustness of the GCL-diversity tradeoff to a different measure of diversity - $\boldsymbol{N_{\mathbf{effF}}}$, which represents the number of expressed genes that contributes to 90% of the total transcription. Across tissues and cell types scRNA-seq data analysis of GCL and $\boldsymbol{N_{\mathbf{effF}}}$ of different organisms. The dots represent different cell type cohorts of single cell genes expression data, collected from different organisms. The same cohorts were analyzed as in Fig. 2. **a,** Pearson correlation rank is -0.12 ($P_{val}$ > 0.2), Z-score=1.62. **b,c,d,** Human gut cross sectional samples from: healthy adults, pediatric and fetal donors. Z-scores of 1.3, 2.59 and 3.04, and Pearson correlation rank of -0.24 ($P_{val}$<0.05), -0.14 ($P_{val}$<0.12) and -0.24 ($P_{val}$<0.005) respectively. **e,** Mice cross tissues samples Pearson correlation rank is -0.23 ($P_{val}$ < 0.005) and Z-score=3.23. In **f.,** Fruit-fly cross tissue samples were analyzed, Pearson correlation rank is -0.21 ($P_{val}$ < 0.07) and Z-score=3.99. **g.,** C-elegans cross tissue samples were analyzed, Pearson correlation rank is -0.29 ($P_{val}$ < 0.11) and Z-score=1.44. Overall, most of the analyzed data exhibit significant GCL-diversity tradeoff pattern where in the majority of the datasets the correlation is significantly negative, and the Z-score>2.

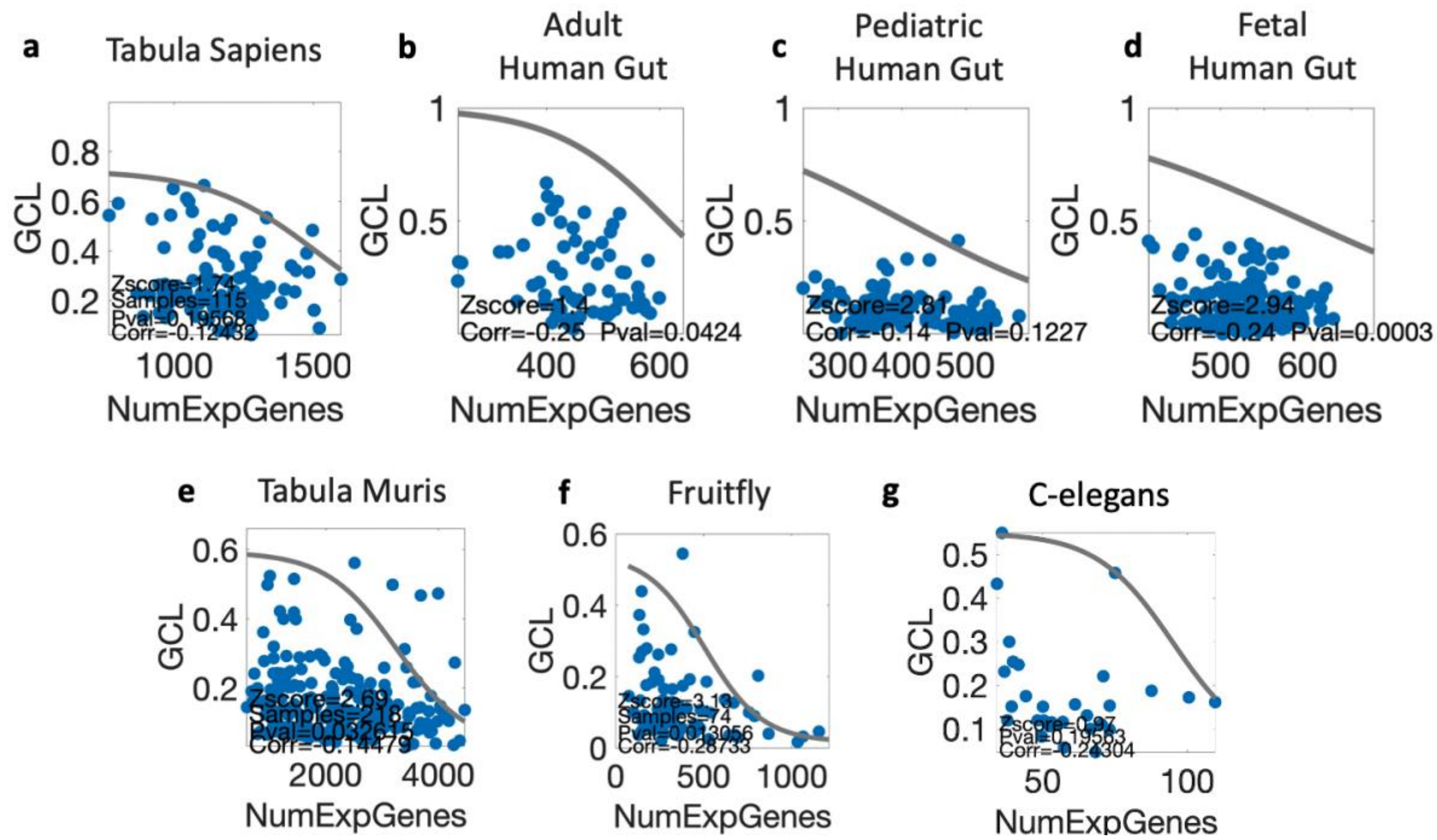

**SI Fig. 8 | $\boldsymbol{GCL\text{-}Number\ Of\ Expressed\ Genes}$ trade-off observed in real scRNA-seq data.** Here we present the robustness of the GCL-diversity tradeoff to a different measure of diversity - $\boldsymbol{N_{ExpGenes}}$, which represents the number of expressed genes in samples with constant number of counts (see Methods). Across tissues and cell types scRNA-seq data analysis of GCL and $\boldsymbol{N_{ExpGenes}}$ of different organisms. The dots represent different cell type cohorts of single cell genes expression data, collected from different organisms. The same cohorts were analyzed as in Fig. 2. **a,** Pearson correlation rank is -0.19 ($P_{val}$ < 0.12), Z-score=1.74. **b,c,d,** Human gut cross sectional samples from: healthy adults, pediatric and fetal donors. Z-scores of 1.4, 2.81 and 2.94, and Pearson correlation rank of -0.25 ($P_{val}$<0.05), -0.14 ($P_{val}$<0.12) and -0.24 ($P_{val}$<0.005) respectively. **e,** Mice cross tissues samples Pearson correlation rank is -0.14 ($P_{val}$ < 0.05) and Z-score=2.69. In **f.,** Fruit-fly cross tissue samples were analyzed, Pearson correlation rank is -0.28 ($P_{val}$ < 0.02) and Z-score=3.13. **g.,** C-elegans cross tissue samples were analyzed, Pearson correlation rank is -0.24 ($P_{val}$ < 0.19) and Z-score=0.97. Overall, most of the analyzed data exhibit significant GCL-diversity tradeoff pattern where in the majority of the datasets the correlation is significantly negative, and the Z-score>2.

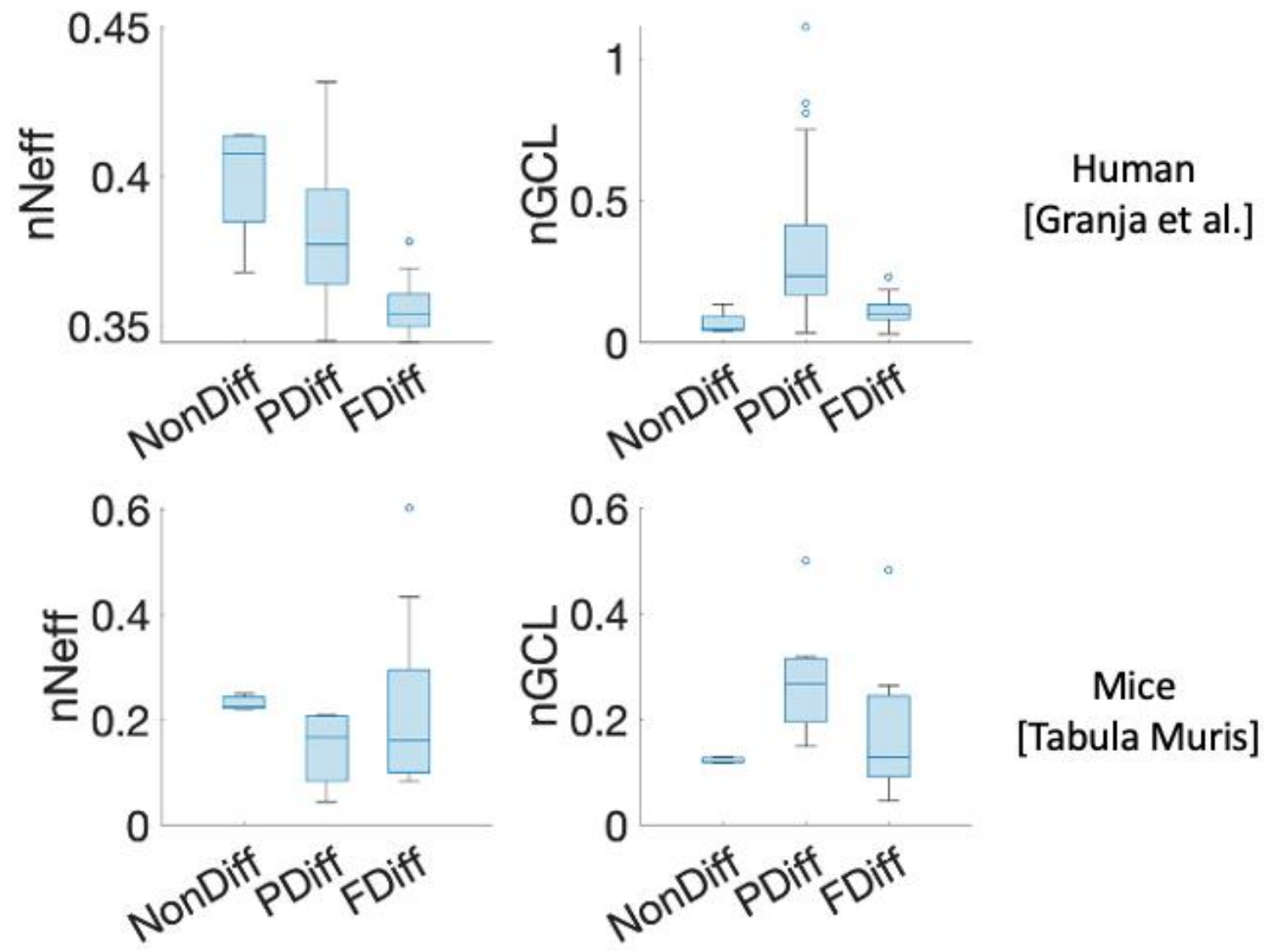

**SI Fig. 9 | GRN intensity – diversity tradeoff patterns upon differentiation.** Here we present the normalized values of $N_{\text{eff}}$ (normalized by the maximal $N_{\text{eff}}$ value of the different cell types) and GCL (normalized by the mean GCL value of the different cell types) of the three groups of differentiated HSCs from Fig. 3a. Similar patterns appears in both human and mice, where the $N_{\text{eff}}$ decreases during differentiation, and the GCL, as the overall complexity (Fig. 4a) is high in transition states partially differentiated cells.

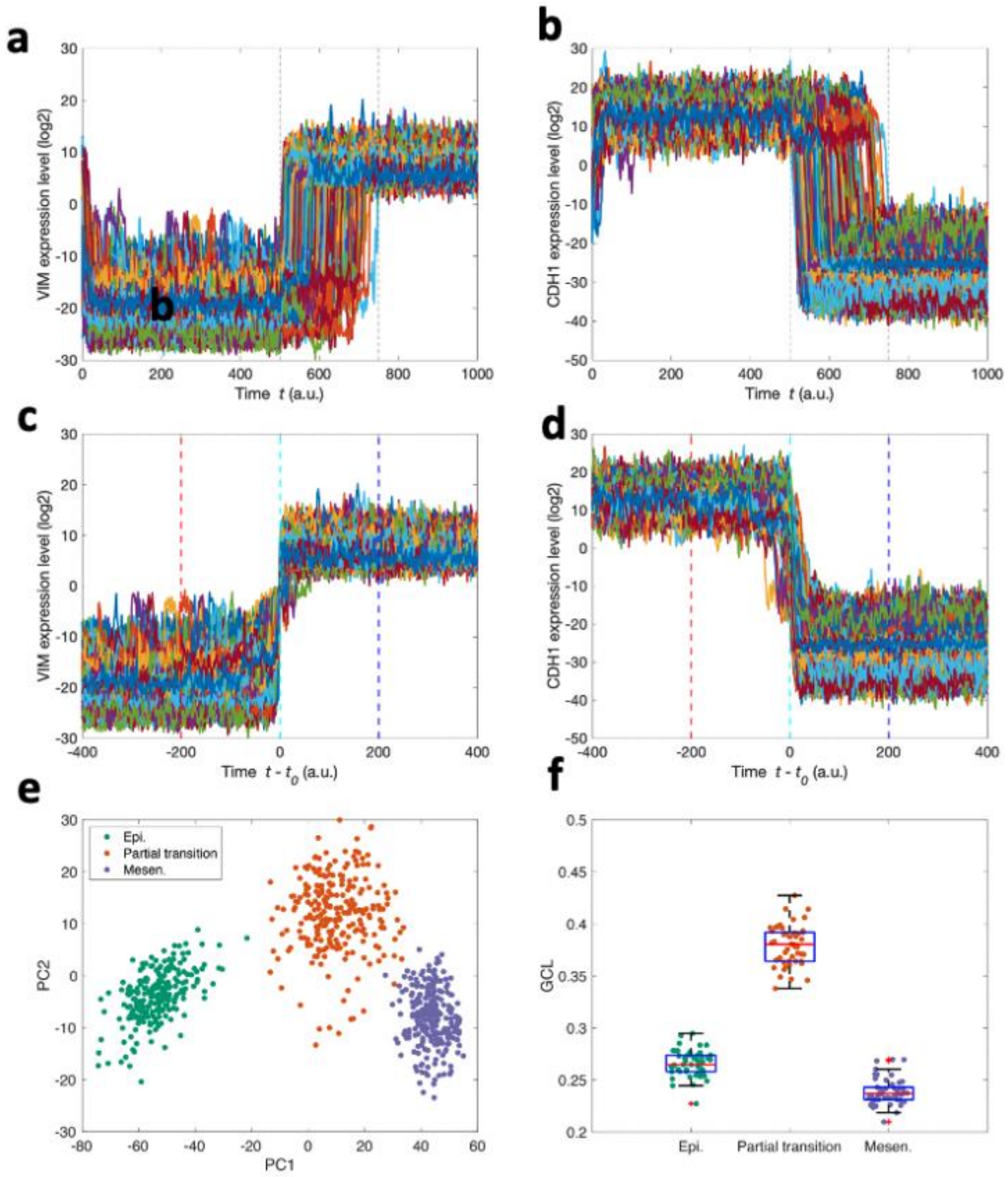

**SI Fig. 10 | Change in GCL during a simulated cell-fate transition**. Here, we used a previously described gene regulatory network involved in regulating the choice between epithelial and mesenchymal cell fates (PMID: 28362798) to simulate transition of cells from epithelial to mesenchymal phenotype. (A) Expression levels of VIM, a key mesenchymal marker, and (B) of CDH1, a key epithelial marker during the simulated cell-fate transition. Each trajectory corresponds to a distinct simulation run and can be interpreted as describing the response of a single cell. Between t = 500 and t = 750 (interval indicated by the black, dashed lines in panels A and B), the expression of two key transcription factors, SNAI1 and ZEB1, is upregulated to induce a transition from epithelial to mesenchymal state. Note that the transition from epithelial to mesenchymal state occurs at different time points in the different trajectories. Only trajectories that exhibit an epithelial expression pattern (low VIM, high CDH1) between t = 200 and t = 500 and transition to a mesenchymal state (high VIM, low CDH1) between t = 500 and t = 750 are shown. These trajectories were used for all subsequent analysis. (C) and (D) show the same trajectories as panels A and B with shifted time: the trajectories are aligned such that all cells undergo a transition around shifted time point 0. We sampled the population of cells at three points: (1) at the time point indicated by the dashed red line ($t - t_0 = -200$) when all cells are in the epithelial state (low VIM, high CDH1; Epi), (2) at the transition time point indicated by the dashed cyan line ($t = t_0$) when the cells have partially transitioned (partial transition), and (3) at the time point indicated by the dashed blue line ($t - t_0 = 200$) when all cells have transitioned to a mesenchymal state (high VIM, low CDH1; Mesen.). (E) Principal component projection of the expression levels of various genes

in the cells sampled at the three points. The figure indicates that the partially transitioned cells have distinct gene expression compared to epithelial and mesenchymal cells, and that this population is not simply a mixture of epithelial and mesenchymal cells. (F) GCL for the populations of cells in epithelial (Epi.), partially transitioned (Partial transition), and mesenchymal (Mesen.) states. Consistent with the behavior reported in Fig. 4, the GCL first increases and then decreases during the simulated cell-fate transition process shown here.

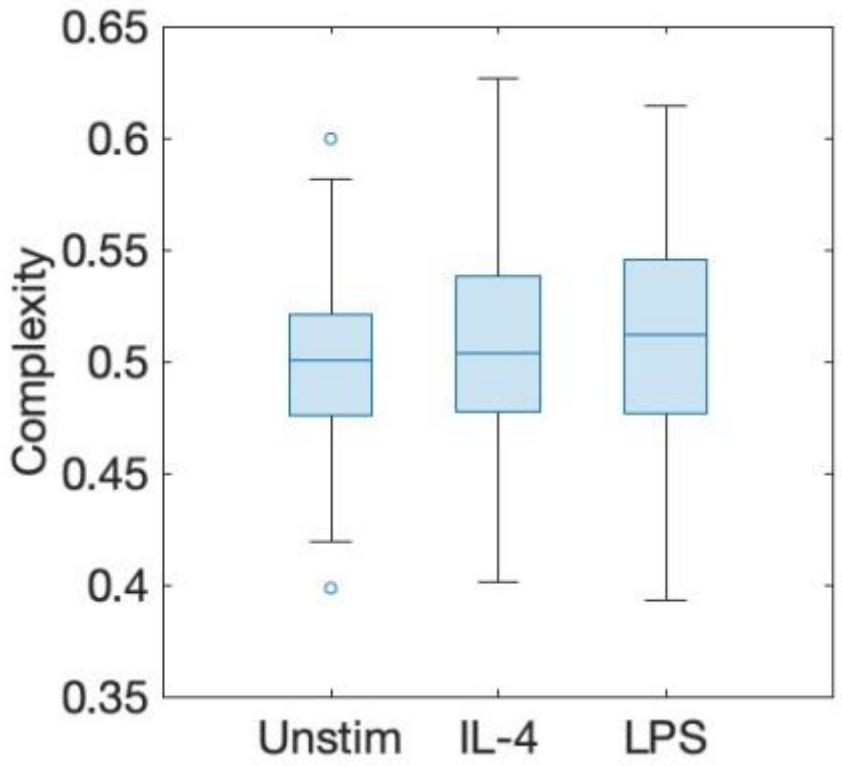

**SI Fig. 11 | Complexity changes upon stimulation**. Here we present the complexity analysis which is based on the normalized $N_{\text{eff}}$ and GCL values of Macrophages unstimulated, LPS and IL4 stimulated cell cohorts. A complementing analysis to the observed GCL and $N_{\text{eff}}$ tradeoff (Fig. 4b), we also see that the complexity remains relatively constant.

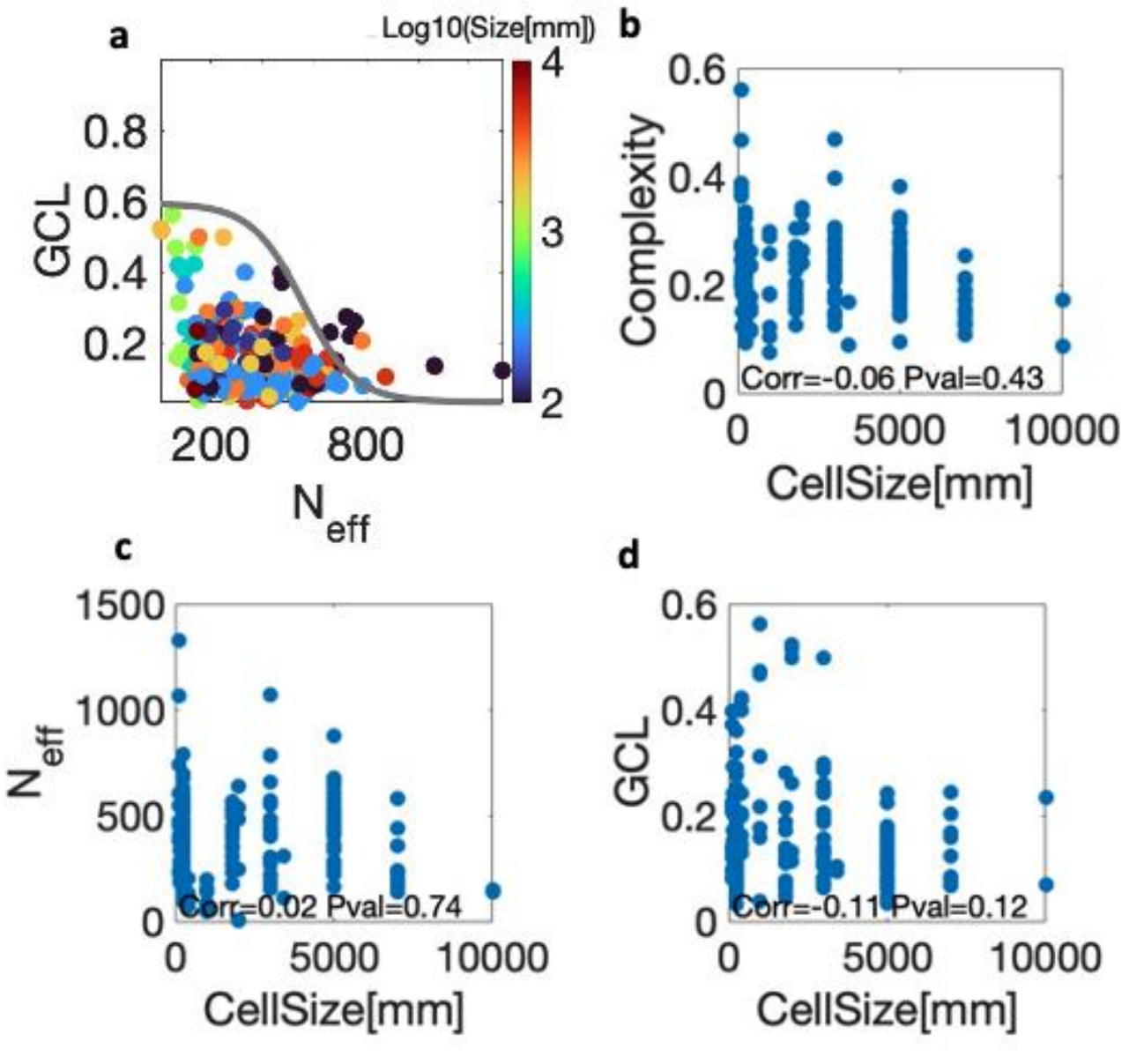

**SI Fig. 12 | Cell size as a function of GRN complexity as measured with GCL and $N_{\text{eff}}$.** Here we present the sizes of different cell types as a function of complexity measures in Tabula Muris dataset [32]. **a,** GCL-$N_{\text{eff}}$ tradeoff, **b,** Complexity**, c,** $N_{\text{eff}}$ and **d,** GCL. No significant correlation patterns are shown.

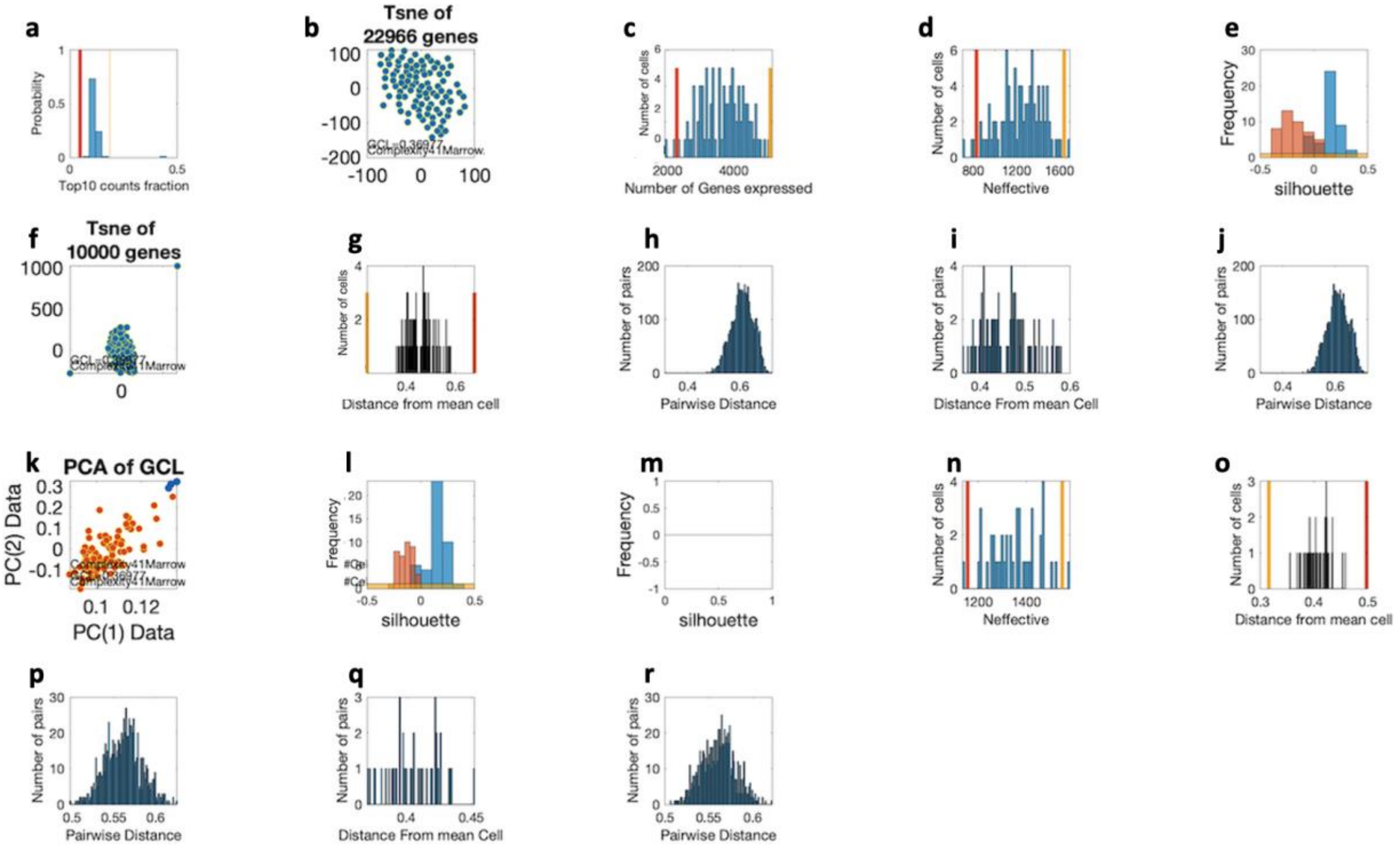

**SI Fig. 13 | Cell filtering process demonstrated for young mouse LT-HSCs [32]. a,** Fraction of the counts sum of the top 10 most highly expressed genes from the total counts in the cell, the red and yellow vertical lines represent the two STDs of the distribution. **b,** TSNE of all cells from the original dataset including all genes. **c,** Number of expressed genes distribution, the red and yellow vertical lines represent the two STDs of the distribution. **d,** Effective number of expressed genes ($N_{\mathrm{eff}}$, as defined in Methods) is larger or smaller than two standard deviations from the mean $N_{\mathrm{eff}}$. **e,** Silhouette analysis of k-means algorithm compared with the realization of the Silhouette analysis of a null model, where the cells are randomly assigned into groups of the same size as the k-means realization. Here we see a significant separation between the two distributions suggesting a significant cluster. **f**, TSNE of all cells after the first step of cell filtering including all genes. **g,** Histogram of the Spearman distances from the average cell. **h,** Pairwise Spearman distance of all cells. **i,** Histogram of the Spearman distances from the average cell and **j,** Pairwise Spearman distance of all cells, both **i** and **j** are plotted after the removal of cells with distance smaller than two standard deviations from the average distance and after the removal of the significantly similar cells. **k,** Final clusters removal that were revealed only after the removal of outliers, we repeat the k-means approach over the PC1-PC2 plane from PCA. Cells that we removed are colored in red and cells that we kept are colored blue. **l,m**, Silhouette analysis of k-means algorithm compared with the realization of the Silhouette analysis of a null model, where the cells are randomly assigned into groups of the same size as the k-means realization. In **l** we see a significant separation between the two distributions suggesting a significant cluster, in **m,** no clusters were found. **n,o,p,q,r,** repeating steps **d,g,h,i,j** respectively.

| Organism | Tissues | Study |
|---|---|---|
| Human | All | Tabula Sapiens [31] |
| Human | Gut | Human Intestinal [33] |
| Mice | All | Tabula Muris [32] |
| C-Elegans | All | C. elegans age [35] |
| Fruit-Fly | All | Fruit-Fly Cell Atlas [34] |
| Human | HSCs and PBMCs | Single-cell multiomic analysis identifies regulatory programs in mixed-phenotype acute leukemia [36] |
| In-Vitro | Macrophages | |

**Extended Data Table 1 | List of scRNA-seq data sets analyzed in this work.** All the data sets analyzed in this work have been published, asides from the stimulation data. The original experiments and corresponding analysis have been reported in previous publications.

| Mice cell types | Human cell types | |
|---|---|---|
| 1 - HSC | 1 - HSC | 2 - CD16.Mono |
| 2 - precursor B cell | 2 - Early.Eryth | 2 - Unk |
| 2 - immature B cell | 2 - CMP.LMPP | 2 - CLP.2 |
| 2 – promonocyte | 2 - LP.1 | 2 - Pre.B |
| 3 - Granulocyte | 2 - GMP | 2 - B |
| 3 - naive B cell | 2 - GMP.Neut | 3 - CD8.N |
| | 2 - pDC | 3 - CD4.N1 |
| | 2 - cDC | 3 - CD4.M |
| | 2 - CD14.Mono.1 | 3 - CD8.CM |
| | 2 - CD14.Mono.2 | 3 - NK |

**Extended Data Table 2 | List of scRNA-seq data sets analyzed in this work.** All the data sets analyzed in this work have been published, asides from the stimulation data. The original experiments and corresponding analysis have been reported in previous publications.